\documentclass[%
 reprint,
 amsmath,amssymb,
 aps,
nofootinbib]{revtex4-2}

\usepackage{float}
\usepackage{amsmath,latexsym,amssymb,amsfonts}

\usepackage{dcolumn}% Align table columns on decimal point
\usepackage{appendix}
\usepackage{mathtools}
\usepackage{graphicx}

\usepackage{parallel,enumitem}
\usepackage[usenames,dvipsnames]{color}

\usepackage{dcolumn}% Align table columns on decimal point
\usepackage{bm}% bold math
\usepackage{hyperref}
\usepackage{mathrsfs}
\hypersetup{
    colorlinks=true,
    linkcolor=blue,
    filecolor=magenta,      
    urlcolor=cyan,
    pdftitle={Overleaf Example},
    pdfpagemode=FullScreen,
    }
\usepackage{tensor}
\usepackage{hyperref} % Add this in the preamble

\begin{document}	
%\title{\textbf{More Nonlinearities? \\ Mode Mixing in NSBH Mergers: Electromagnetic-Gravitational Case}}
\title{\textbf{More Nonlinearities? \\ Electromagnetic and Gravitational Mode Mixing in NSBH Mergers}}

\author{Fawzi Aly${}^{1}$
\href{https://orcid.org/0009-0009-4183-4531}{\includegraphics[scale=0.015]{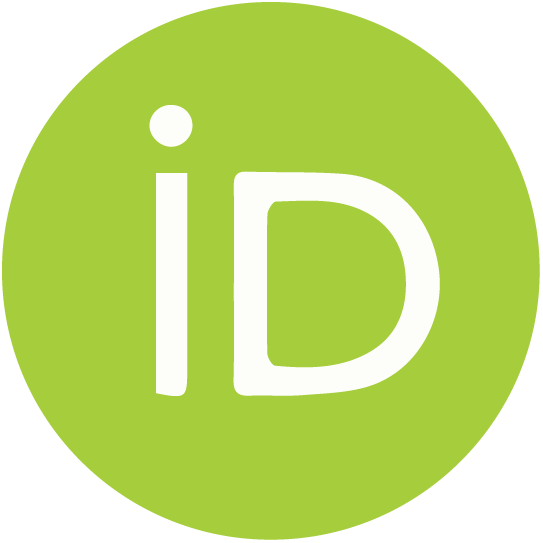}}}
\thanks{Corresponding author}
\email{mabbasal[AT]buffalo.edu}
\author{ Mahmoud A. Mansour${}^2$
\href{https://orcid.org/0009-0005-9634-1700}{\includegraphics[scale=0.015]{figures/orcid.png}}}
\email{ mansour[AT]iis.u-tokyo.ac.jp}
\author{ Dejan Stojkovic${}^1$
\href{https://orcid.org/0000-0002-6894-0539}{\includegraphics[scale=0.015]{figures/orcid.png}}}
\email{ds77[AT]buffalo.edu}
\affiliation{${}^1$HEPCOS, Physics Department, SUNY at Buffalo, Buffalo, New York, USA}
\affiliation{${}^2$Department of Physics, The University of Tokyo, Kashiwanoha, Kashiwa, Chiba 277-8574, Japan}

\begin{abstract}
We investigate the possibility of electromagnetic fields leaving imprints on gravitational wave (GW) signals from Neutron Star-Black hole (NSBH) mergers, specifically in the context of extreme mass ratio inspirals (EMRIs). Using black hole perturbation theory (BHPT) in the context of a minimally coupled Einstein-Maxwell system, we demonstrate that electromagnetic quasi normal modes
(QNMs) can excite gravitational QNMs with frequencies that are  linear or quadratic in the electromagnetic QNMs, at first level of mixing. Moreover, We then study the electromagnetism-gravity coupling by approximating the Regge-Wheeler and Zerilli potentials with Dirac delta functions. In this example, we examine gravitational perturbations induced by the electromagnetic field of an ideal dipole radially free fall towards the blackhole, building on calculations from a companion paper \cite{Fawzi_EM_idealdipole}. Our results show that both linear and quadratic electromagnetic QNMs appear in gravitational perturbations. In addition, linear gravitational QNMs are also excited due to the electromagnetic source, with their amplitudes depending on the details of the electromagnetic and gravitational potentials, analogous to gravitational mode mixing analysis. Furthermore, at late stages, gravitational perturbations might exhibit polynomial tails induced by electromagnetic perturbations. This article sets the stage for future numerical investigations aimed at identifying such modes in various scenarios.
\end{abstract}
\maketitle

\section*{Introduction}
After LIGO's first detection of gravitational waves (GWs) from a stellar-mass black hole binary (BBH) merger in 2015, during the first run, O1, with the landmark event \textbf{GW150914}, involving two black holes of approximately 36 and 29 \(M_{\odot}\) \cite{GW150914}, gravitational wave astronomy took a significant leap forward. In 2020, during O3, LIGO and Virgo further advanced the field with the first confident detection of a neutron star-blackhole (NSBH) binary merger \cite{GW200105_GW200115}. The events \textbf{GW200105} and \textbf{GW200115} involved black holes of 8.9 and 5.7 \(M_{\odot}\) merging with neutron stars of 1.9 and 1.5 \(M_{\odot}\), respectively. With ongoing improvements in detector sensitivity, the number of NSBH detections is expected to rise in the coming decade. However, NSBH mergers are still expected to occur less frequently than BBH or even binary neutron star (BNS) mergers \cite{Gupta, rate_NSBH_merger}.
\vspace{1mm}

\noindent LISA will extend the observational capabilities of GW astronomy. Unlike LIGO, which observes higher-frequency GW in kHz band from stellar-mass systems, LISA will be sensitive to gravitational waves in the $10^{-3}$ - $10^0$ Hz range, focusing more on the inspiral phase of mergers. That makes it ideal for detecting systems with total masses between \(10^4 \) - \(10^7\) \(M_{\odot}\) \cite{Aly_2023,TestingGravityWithEMRIs,Amaro_Seoane_2023,Berti_2006}. A key example of such systems is Extreme Mass Ratio Inspirals (EMRIs), where a compact object, such as a NS spirals then latter plunge into a supermassive black hole \cite{Gair_2017_Detecting_EMRI_LISA,EMRI_SNR_2004_Gair}. EMRIs are modeled using black hole perturbation theory (BHPT) within the self-force program \cite{Pound_2021}, and LISA is expected to probe these systems with high signal-to-noise ratios (SNR) $>$ 50 \cite{amaroseoane2017LISA}. Although the exact detection rate of EMRIs remains uncertain, it is estimated that LISA could detect between $10^1$ - $10^3$ events per year \cite{rate_EMRI_1,rate_EMRI_2,Amaro_Seoane_2023}.
\vspace{0.5mm}

\noindent With the advancement of LIGO's GW detection, there has been a parallel shift in fitting models which originally took into account only the fundamental gravitational mode $(l, m, n) = (2, 2, 0)$. Over last few years, researchers took into account higher numbers of overtones $n>0$ and/or higher harmonics $(l,m)$\footnote{Higher harmonics should be relevant also for waveforms with $q>1$, specifically SMBHBs \cite{pitte2023detectabilityhigherharmonics}.}\cite{OverTonesImportance_Giesler,Isi_testing_No_Hair_Theorem,OverTone8,2Modes_As_test_Per_theory_1,HigherHarmonics,HigherOvertones_Bhagwat,Overtone_lessEvidentbutstill_Finch}. Alternatively, one can include nonlinearities \cite{NonlinearitiesCaltechMitman,Cheung_Nonlinearities_Berti_Grouup,NearHorizonNonlinearities}, suggesting that nonlinear models may provide a more accurate fit to ringdown waveforms compared to former linear ones. This trend is evident in numerical templates for astrophysical mergers and also simulations of head-on collisions, in waveforms both in the  asymptotic spacetime regime and in the near-horizon regions where nonlinear effects are expected to be more pronounced \cite{NearHorizonNonlinearities}. 
\vspace{.5mm}

\noindent On the other hand, NSBH and BNS systems exhibit greater dynamical intricacy compared to BBH systems leading to a richer phenomenology, which makes them particularly interesting candidates for multi-messenger astronomy\cite{multi_messenger}. One of the distinctive features of these systems is the role of electromagnetic interactions. Subsequently, a natural question arises: \textit{“Is there more to consider when modelling GW signals from NSBH/BNS mergers?”}
\vspace{1mm}

\noindent One potential factor of interest is whether the highly magnetized nature of NS could influence the detected GW signal through their electromagnetic fields. This could be studied in the framework of BHPT\footnote{Whether formulated in the metric-perturbation approach \cite{KerrMetricPerubation_Dolan_2024, spiers2024_secondorder_perturbations_schwarzschildspacetime, KarlMartel_2005_RW, Brizuela_2006_metricperubationhighorderSchwarzschild, Brizuela_2009_metricperubationhighorderSchwarzschild, Brizuela_2010, Ioka_SecondSchwarzschild, chandrasekhar_1983} or in the curvature-perturbation approach \cite{chandrasekhar_1983,KerrNPFormalism_Loutrel,Chitre_ChargedBH_pertubation,spiers2023_second_teukolsky_kerr}}, where the electromagnetic coupling will, as shown in this work, contribute to the QNMs spectrum. Such change could be examined perturbatively, analogous to gravitational mode-mixing studied in the literature which are referred to here as \textit{GG modes} \cite{Gleiser_Second_Schwarzschild1996,Gleiser_Second_Schwarzschild1996_CloseApprox,Gleiser_Second_Schwarzschild2000,Lagos_2023_GreenFunctionAnalysis_Quadratic_Diracdelta,Okuzumi_Ioka_2008,Ioka_2007,Ioka_SecondSchwarzschild}. The GG modes arise when gravitational modes, refereed to as \textit{G} modes in this article, serve as an effective source term which drives the higher-order gravitational perturbations. Simply enough, matter sources would give rise to electromagnetic-gravitational mode mixing, in the same manner, when first-orders electromagnetic mode \textit{EM} and G mode mix.
\vspace{1mm}

\noindent Aside from the intriguing theoretical aspects of this research endeavor, more importantly, it remains unclear at which order the electromagnetic-gravitational coupling becomes significant, potentially leaving detectable imprints in the signal.
\vspace{1mm}

\noindent Usually, in EMRIs \cite{Pound_EMRI_SF_intro, EMRI_ScalarClouds, Poisson_Pound_EMRI}, matter is modeled as either a point particle or tightly localized distribution following a geodesic on the unperturbed spacetime, where the center of mass frame is approximately the super massive black hole (SMBH) rest frame. In that sense, the mass ratio $m/M$ and the ratio between the electromagnetic energy and the mass of the supermassive black hole $M$, $U_{\text{EM}}/Mc^2$, could serve as small perturbation parameters for the coupled gravity-electromagnetism system.
\vspace{1mm}

\noindent It’s conceivable to imagine a NS or magnetar of mass $m$ somewhere between $1-2$ $M_\odot$ and radius $R$ of around 10 km, with a magnetic field strength ranging from \(10^{7}\) - \(10^{11}\) \textbf{T}. If most of the magnetic field extends beyond the neutron star into a region approximately $1-2$ times its radius, a rough estimate of the energy content in the surrounding area could be
\[
U_B \approx 10^{34} - 10^{42} \textbf{J}.
\]
When compared to a SMBH of mass $M$ measured in units of $M_\odot$, the energy ratio is 
\[
\frac{U_B}{Mc^2}=\frac{10^{-13} M_\odot}{M} \ \ {\rm to} \ \ \frac{10^{-5} M_\odot}{M}.
\]
Meanwhile, for a NS of two solar masses, i.e. $m=2 M_\odot$, the EMRI mass ratio is
\[
\frac{m}{M} \approx \frac{2 M_\odot}{M}.
\]
By comparing these two small parameters, if we want the lower limit of this ratio  to be significant in the second-order perturbation theory (which is a conservative assumption), $M$ would need to be of the order of $10^6$ $M_\odot$, for that the lower bound of electromagnetic field strength to start compete with second order correction for the mass $m$ of the perturber
\[
 \left(\frac{m}{M}\right)^2 \approx 10^{-12}\quad\quad \frac{U_B/c^2}{M}>10^{-13},
\]
although more concrete calculation are warranted. This plausibly suggests that the effects of gravitational and electromagnetic mode mixing might be relevant for a merger of a NS with a SMBH.
\vspace{.5mm}
%Paragraph, in stellar mass BNS merger such as \textbf{GW170817},
 
%Paragraph about QNMs in numerical simulation paper as another motivation

%Commonly, plasma or accretion disks surrounding a remnant black hole can leave fingerprints in its spectrum as well, typically studied in magnetohydrodynamic simulations of black holes. However, strong electromagnetic radiation during the merger phase can more easily leave detectable fingerprints in the signals due to its discrete spectrum.

\noindent Furthermore, it is not only the perturbation order that dictates how \textit{G} modes and \textit{EM} modes mix; the coupling mechanism also plays a significant role in shaping the resulting QNM spectrum. Therefore, if such \textit{GEM} modes could plausibly be detected, they could potentially serve as a test for minimal coupling within the framework of General Relativity(GR), as well as for exploring alternative couplings in modified theories of gravity.
\vspace{1mm}

\noindent In this work, we will study the Einstein-Maxwell system perturbatively and formulate how the EM modes and G modes couple to generate the GEM modes. We will employ the Laplace integral method to demonstrate that linear and quadratic EM modes appear as poles in complex frequency space. Next, the inverse Laplace transformation will be used to solve for the second order gravitational perturbation, in a similar fashion to GG modes. Furthermore, we will analyze a toy model where an ideal dipole perturber of mass $2m$ and dipole moment $p=q \eta$ perturbs the Schwarzschild black hole spacetime of mass $M$. To simplify the problem, we replace the actual potentials with Dirac delta functions for both the gravitational and electromagnetic perturbation equations. In this example, we will illustrate how GEM modes manifest.
\vspace{1mm} 
%For instance, considering GW200105 and GW200115 observations, and given the strong magnetic fields of neutron stars, it raises the question: \textit{Could electromagnetic radiation coupling significantly influence the modeling of linear/nonlinear regimes of the ringdown of the remnant black hole?} The black hole remnant in those observations, as well as similar cases, is expected to be born with magnetic hair. However, according to the no-hair theorem and the uniqueness theorem, the black hole, if isolated, should quickly settle into the Kerr family spacetime and lose its magnetic hair. This is typically believed to happen through radiation during the merger/post-merger phases. Even in more realistic scenarios where the binary is surrounded by dust such as plasma, recent work has shown that the black hole will still decay its magnetic hair exponentially, although we are not aiming to tackle this latter case in this write-up. Equivalently, our latter question translates to: \textit{How violently does a black hole lose its electromagnetic hair?}
%\vspace{1mm}

%In principle, in astrophysical scenarios of NSBH mergers where no disruption took place, some of the magnetic field will be inherent in the newborn black hole. Consequently, questioning the influence of the electromagnetic-gravity coupling at linear and quadratic QNM frequencies in the perturbation scheme is a legitimate inquiry, especially since this type of EM-G mode mixing has different features from the normal G modes and G-G modes.
%\vspace{1mm}

\noindent The structure of this paper is as follows. In subsection \ref{Problem Statement}, we briefly outline the technical aspects of the problem under the assumption of minimal coupling, followed by a comparison of the characteristics of GG and GEM modes. In subsection \ref{StressEnergyTensor_Recasting}, we describe the recasting process of the electromagnetic stress-energy tensor $T_{EM\, \mu \nu}$, expressed in terms of the electromagnetic perturbation scalars $ {}_{o}\Phi_{lm} $ and $ {}_{e}\Phi_{lm} $, to illustrate how electromagnetic parents modes with angular numbers $(l_1,m_1)$ and $(l_2,m_2)$ mix to induce gravitational mode with $(l,m)$. Detailed analysis can be found in appendices \ref{Decomposition Using Spherical Harmonics in Schwarzschild Spacetime}, \ref{Useful Spherical Integrals}, \ref{Electromagnetic Source Terms for the Regge-Wheeler and Zerilli-Moncrief Equations}, and \ref{The Spherical Decomposition of the Electromagnetic Energy-Momentum Tensor}. In section \ref{Formulation_Mode_Mixing}, we provide a step-by-step explanation of how electromagnetic QNMs with frequencies $\Omega$ induce gravitational QNMs with frequencies linear and quadratic in $\Omega$, referred to as GEM modes. Readers familiar with GG  analysis in BHPT can proceed directly to subsection \ref{GEM Modes}. Additionally, we illustrate the existence of GEM modes using a Dirac delta function toy model in section \ref{toy_model}, starting with the electromagnetic QNM of an ideal dipole moment, as derived in the companion paper \cite{Fawzi_EM_idealdipole}, and use this to source the gravitational perturbations in subsection \ref{Gravtional IDeal dipole}. Finally, we discuss the results of our analysis and potential directions for future work in section\ref{Conclusion}. Throughout this work, we use natural units where $G = c = 1$.

\section{Preliminary}
In essence, a QNM of a black hole spacetime is a characteristic pair of a frequency and a decay time of the  mode describing excitations of the system. They are determined by the black hole’s macroscopic properties, primarily\footnote{This assumes that the remnant is formed without a macroscopic charge. However, considering charged black hole could be important too, especially in the context of hidden U(1) fields \cite{Aly_2024}.} its mass M and spin J. Observing a single QNM provides a measurement of these parameters, while detecting two or more QNMs serves as a consistency check, testing the accuracy of BHPT in GR, which is expected to accurately model the late-stage ringdown signal \cite{2Modes_As_test_Per_theory_1,2Modes_As_test_Per_theory_2_Agnosticspectroscopy}. However, when the black hole spacetime is governed by more than just the Einstein field equations, e.g. when both the Einstein's and Maxwell's equations are involved\footnote{More generically, spin-2 coupled to spin-1 theory}, the QNMs expand to include the GEM modes. In the following subsection \ref{Problem Statement}, we outline the technical aspects of the problem, assuming minimal coupling. Additionally, we provide a comparison between the characteristics of GG and GEM modes. Next, in subsection \ref{StressEnergyTensor_Recasting}, we outline the recasting of the electromagnetic stress energy tensor $T_{EM\, \mu \nu}$ written in terms of the electromagnetic perturbation scalars ${}_{o}\Phi_{lm}$ and  ${}_{e}\Phi_{lm}$ to adjust it for separation of the angular dependence in the gravitational perturbation as outlined in \cite{KarlMartel_2005_RW}. 

\subsection{Problem Statement}\label{Problem Statement}

\begin{figure}
\includegraphics[width=8cm]{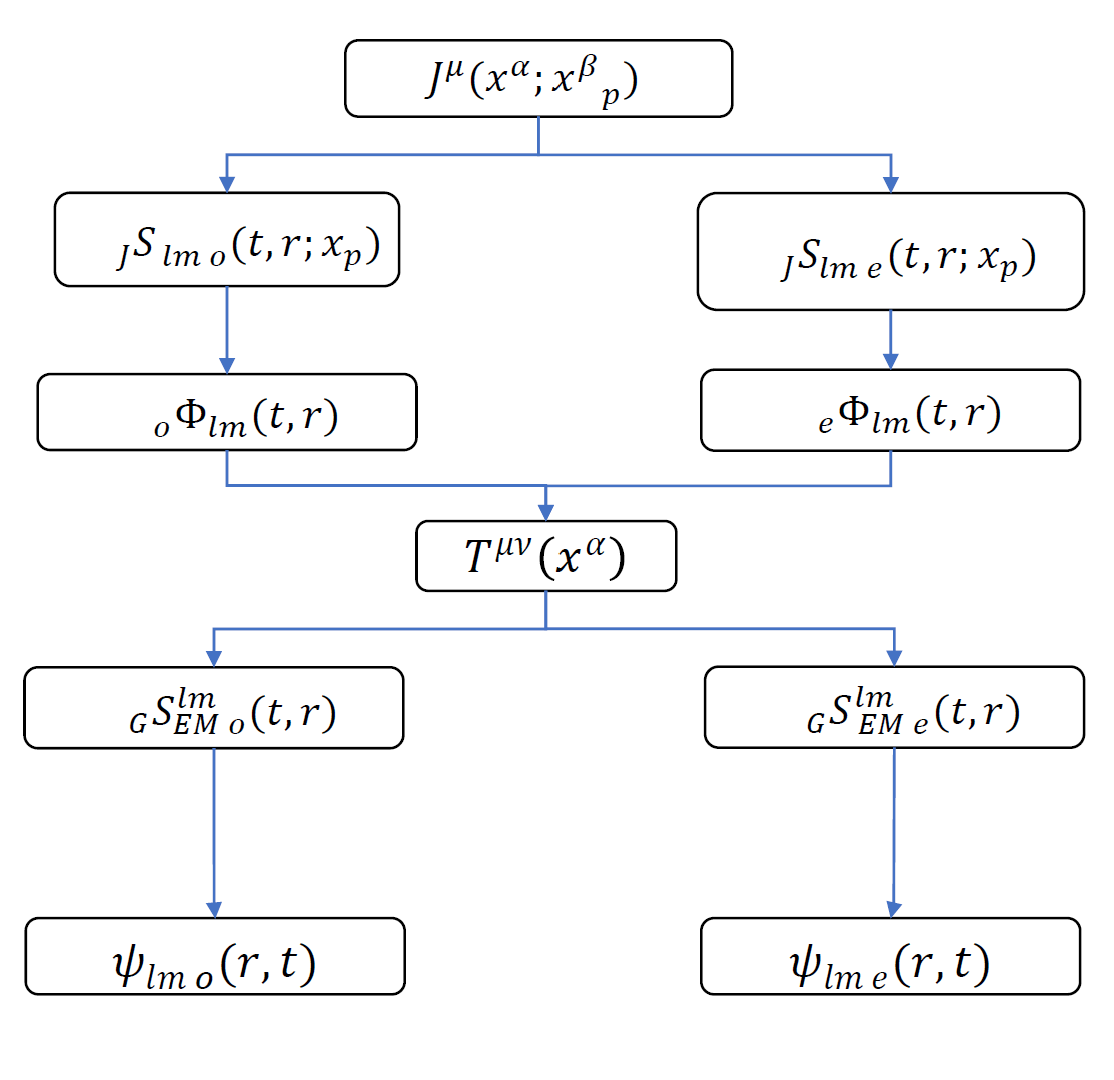}
\caption{This diagram illustrates the iterative solution of the perturbed Einstein-Maxwell equations, assuming subdominant EM effects. Starting with the EM source $J^{\mu}$, we solve for the EM perturbation $\Phi$ and compute the EM stress-energy tensor $T^{\mu\nu}_{EM}$. This acts as a source for gravitational perturbations, producing GEM modes ${{}_{EM}\psi_{G}}$. If $T^{\mu\nu}_{EM}$ and the gravitational stress-energy tensor $T^{\mu\nu}_{GG}$ are comparable at some order $k$, the total gravitational perturbation $\psi$ will be a combination of GEM modes ${{}_{G}\psi_{EM}}$ and GG modes ${{}_{G}\psi_{GG}}$.}
\label{generalmap}
\end{figure}

Assuming that the remnant (or the SMBH in the case of EMRIs) is neutral in charge and acquires its spin, $J$, from the angular momentum of the progenitors and their individual spins, our problem can be treated as a Kerr perturbation problem. However, for simplicity in this article, we will focus on the Schwarzschild background as a proof of concept. We believe the same analysis can be extended to the Kerr case using the Newman-Penrose formalism, as illustrated in figure \ref{generalmap}. 
\vspace{1mm}

\noindent The perturbed Einstein-Maxwell field equations are solved iteratively, under the assumption that the EM effects are subdominant and only appear in higher orders of gravitational perturbations. Starting with the EM source, $J^{\mu}$, we can solve for the EM perturbation, $\Phi$, and then calculate the corresponding EM stress-energy tensor, $T^{\mu\nu}_{EM}$. On the gravitational side, this tensor will act as a source for the gravitational perturbation equation, allowing us to solve for the GEM modes, ${{}_{G}\psi_{EM}}$. Given the linearity of the master perturbation equations, if the EM stress-energy tensor, $T^{\mu\nu}_{EM}$, and the gravitational self-interaction effective stress-energy tensor, $T^{\mu\nu}_{GG} $, are of comparable magnitude at some order $k$, the gravitational perturbation, $\psi$, will be a linear combination of both GEM, ${{}_{G}\psi_{EM}}$, and GG modes, ${{}_{G}\psi_{GG}}$.
\vspace{1mm}

\begin{figure}
\includegraphics[width=8.5cm]{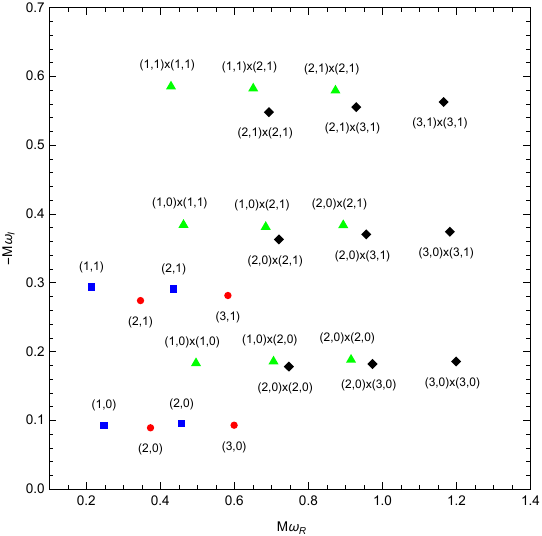}
\caption{Comparison of G and EM QNMs in Schwarzschild spacetime. G frequencies $\omega_{ln}$ for $l = 2, 3$ and $n = 0, 1$ are compared with EM frequencies $\Omega_{ln}$ for $l = 1, 2$ and $n = 0, 1$. For instance, the GG, quadratic QNMs, the real part of the frequencies is given by the sum $\omega_{(i\times j)R}=\omega_{(i)R} \pm \omega_{(j)R}$ of the linear modes, while the imaginary part, representing the reciprocal of the decay time, is the sum $\omega_{(i\times j)I}=\omega_{(i)I} + \omega_{(i)I}$ of the corresponding linear terms. The blue squares and red dot refer to the electromagnetic and gravitational QNMs while green triangles and rhombuses refer to the quadratic electromagnetic and gravitational QNMs respectively. }
\label{fig:myPlot}
\end{figure}

\noindent In figure \ref{fig:myPlot}, we compare the gravitational QNMs in Schwarzschild spacetime with their electromagnetic counterparts. Specifically, we include the gravitational modes for $l = 2, 3$ and $n = 0, 1$, and the electromagnetic modes for $l = 1, 2$ and $n = 0, 1$. The mode $(2,0)$ represents the fundamental QNM for gravitational perturbations, while for electromagnetic perturbations it's the $(1,0)$. Within each subset, the decay time does not vary significantly for a constant overtone number $n$. However, both decay time and frequency increase as the overtone number $n$ and angular momentum number $l$ increase respectively. Additionally, we include quadratic QNMs, which are formed by the outer product of the gravitational and electromagnetic modes, respectively: the quadratic modes such as $(2,0) \times (2,0)$ for gravitational QNMs and $(1,0) \times (1,0)$ for electromagnetic QNMs are expected to appear prominently. With minimal coupling, the linear EM and quadratic EM modes will resemble the GEM modes at some order $k$, when the GEM modes become relevant to the perturbation.

\subsection{Recasting the Electromagnetic Stress-Energy Tensor }\label{StressEnergyTensor_Recasting}
\noindent The spherical decomposition typically applied to the vector potential $A_{\mu}$, as discussed for example in \cite{KarlMartel_2005_RW, Cardoso_EM_Formulation_Anti_de_Sitter}, does not directly generate an electromagnetic stress-energy tensor $T_{EM\, \mu \nu}$ in representation suitable for the Regge-Wheeler spherical decomposition used in separating the gravitational perturbation PDEs to the well known Regge-Wheeler and Zerilli ODEs.\cite{KarlMartel_2005_RW,mARTELphdthesis,Gerlach_Ulrich}. In the appendices \ref{Decomposition Using Spherical Harmonics in Schwarzschild Spacetime}, \ref{Useful Spherical Integrals}, \ref{Electromagnetic Source Terms for the Regge-Wheeler and Zerilli-Moncrief Equations}, and \ref{The Spherical Decomposition of the Electromagnetic Energy-Momentum Tensor}, we recast the electromagnetic stress-energy tensor $T_{EM\, \mu \nu}$ to adapt it for the latter purpose as outlined in \cite{KarlMartel_2005_RW}.
\vspace{1mm}

\noindent Starting with the decomposition of the electromagnetic vector $A_{\mu}$ potential using spherical harmonics and vector spherical harmonics \cite{Multipole_Thorne,edmonds1996angular} %- we can cite Kip thorns paper and a reference book [Thorne, K.S., 1980. Multipole expansions of gravitational radiation. Reviews of Modern Physics, 52(2), p.299, Edmonds, A.R. and Mendlowitz, H., 1958. Angular momentum in quantum mechanics],
as demonstrated in \cite{KarlMartel_2005_RW}, we reformulated Maxwell’s equations similarly to \cite{Cardoso_EM_Formulation_Anti_de_Sitter}. These equations are written in terms of two degrees of freedom corresponding ${}_{e}\Phi_{l'm'}$ and ${}_{o}\Phi_{lm}$ to parities $(-1)^l$ and $(-1)^{l+1}$ respectively, which satisfy the electromagnetic Regge-Wheeler equations sourced by the source ${}_{e} S_{J\, lm}$ and ${}_{o} S_{J\, lm}$ mixing the four-current $J_{\mu}$ components. Using these degrees of freedom and the four-current, we extended the decomposition to the energy-momentum tensor. The explicit calculation of this decomposition, following the methodology of \cite{KarlMartel_2005_RW}, is detailed in the appendices.
\vspace{1mm}

\noindent All integrals involved in the decomposition can be reduced to two main types: one quadratic in the degrees of freedom of the same parity ${}_{e}\Phi_{lm} \times {}_{e}\Phi_{l'm'}$ or ${}_{o}\Phi_{lm} \times {}_{o}\Phi_{l'm'}$ and the other coupling the degrees of freedom of different parities ${}_{e}\Phi_{lm} \times {}_{o}\Phi_{l'm'}$. Each of these integrals reveals a simple coupling between the gravitational modes $(l,m)$ and the electromagnetic $(l_1,m_1)$ and $(l_2,m_2)$ modes. This clear coupling structure makes it straightforward to determine which modes of the electromagnetic field source a particular gravitational mode.

\section{Formulation: Mode Mixing}\label{Formulation_Mode_Mixing}
To account for the effect of the electromagnetic field of e.g. a NS, we need to couple the electromagnetic field degrees of freedom, taken as $A_{\mu}$(x), to the metric functions $g_{\mu \nu}$(x) that describe the gravitational field. This can be achieved through minimal coupling by solving the Einstein-Maxwell system of coupled differential equations
\begin{equation}\label{EM system}
G^{\mu\nu} =8\pi T_{EM}^{\mu \nu}+ 8\pi T_{other}^{\mu \nu} .
\end{equation}
Maxwell equations are sourced by the current four-vector \( J_{\nu} \)
\begin{equation}
\nabla^{\mu} F_{\mu\nu} = J_{\nu},
\end{equation}
where the stress-energy tensor of the electromagnetic field \( T_{EM}^{\mu\nu} \), as well as contributions from other sources \( T_{other}^{\mu \nu} \), including other fields. In this work, we are not including any contributions from additional fields. However, within the perturbative scheme of EMRIs, \( T^{(1)}_{G} \) represents the stress-energy tensor of the perturber mass sourcing $G^{(1)}$, while \( T^{(2)}_{GG} \) and more generally \( T^{(l)}_{GG} \) accounts for the gravitational mode couplings at the second and higher orders \( l \), respectively.  \vspace{1mm}

\noindent The Faraday tensor $F_{\mu\nu}$ sources the gravitational curvature, while the curvature encoded in $g_{\mu\nu}$ influences the evolution of the Maxwell field. Imagine that the four-potential vector \( A \) is of order \(\alpha\) and the full metric \( g \) is of order \(\eta\). The first-order correction of the stress-energy tensor for the electromagnetic field \( T_{EM}^{\mu \nu} \) will be of order \(\alpha^2\) since the background, denoted as \( g^{(0)} \), is uncharged, \( A^{(0)} = 0 \).
\begin{equation}
\begin{aligned}
    F^{(1)}_{\mu\nu} &= \alpha (\partial_{\mu} A^{(1)}_{\nu} - \partial_{\nu} A^{(1)}_{\mu})\\
    T^{(2)}_{EM \, \mu\nu} &= \alpha^2 (F^{(1)}_{\mu\lambda} F^{(1) \lambda}_{\nu} - \frac{1}{4} g^{(0)}_{\mu\nu} F^{(1)}_{\lambda\sigma} F^{(1) \lambda\sigma}).
\end{aligned}
\end{equation}
In the case where the electromagnetic field is switched off, the induced metric $g^{(1)}_{\mu\nu}$ is of the order of \(\beta\) due to the mass of the pertuber. Roughly, a hierarchy\footnote{While it is not quite correct to compare these scales directly, it should provide a rough idea of the system's hierarchy.} is established between \(\alpha\) and \(\beta\) based on the relative strengths of these perturbations, with \(\beta\) generally being dominant in most astrophysical scenarios. However, \(\alpha\) may still be relevant in certain cases, such as in NSBH and BNS mergers, as well as during the collapse of a NS or a Magnetar to a BH.
\vspace{1mm}

\noindent On a Schwarzschild background, we can apply spherical decomposition to the system \cite{KarlMartel_2005_RW}, and write a coupled system of perturbation equations as follows 
\begin{equation}\label{coupled pertubation}
\begin{aligned}
&\mathcal{L}_{G} \, \psi^{(k')} = S^{(k')}_{G}\left[\psi^{(k'-1)},\psi^{(k'-2)},...\right]+ S^{(k')}_{G}\left[T_{EM}\right],\\
&\mathcal{L}_{EM} \,\Phi^{(k)} = S_{EM}^{(k)}\left[J\right],
\end{aligned}
\end{equation}
where \( k, k'\) are the perturbation orders of the electromagnetic field and gravitational fields, respectively. 

\noindent In subsection \ref{BHPT}, we examine the coupled perturbation equations (\ref{coupled pertubation}), following the standard approach commonly found in the literature. Readers already familiar with BHPT may choose to skip this subsection and proceed directly to subsection \ref{GEM Modes}, where we focus on extracting the GEM modes as poles of the gravitational perturbation scalars in the complex frequency domain.

\subsection{BHPT}\label{BHPT}
Klein-Jordan, electromagnetic, and odd gravitational perturbations on Schwarzschild spacetime are described by Schrödinger-like wave equations with different potentials \( {}_{s}V_{RW} \) for each perturbation, where \( s = 0, 1, 2 \) respectively, giving rise to the Regge-Wheeler equation. 
\begin{equation}
{}_{s}V_{RW} = \frac{f(r)}{r^2} \left[l(l+1)+ \frac{2M (1-s^2)}{r} \right],
\end{equation}
where $l$ is the angular momentum number.
Similarly, even gravitational perturbations are governed by a different potential known as \( V_{Z} \), described by the Zerilli equations.
\begin{equation}
V_{Z} = \frac{f(r)}{r^2}\left\{\frac{[(l-1)(l-2)] f(r) r}{2(r+3 M)}+\frac{18 M^3}{(l+1)(l-1)r}\right\},
\end{equation}
although the potentials differ, the differential equations exhibit the same singular structure: same number of poles,  ranks and locations in schwarzschild coordinates. 
\vspace{1mm}

\noindent At order \( k \) of gravitational perturbations, the same differential operator \( {}_{2}\mathcal{L}_{RW} \) (or \( \mathcal{L}_{Z} \)) acts on the perturbation scalar \( {}_{o} \psi^{(k)} \) (or \( {}_{e} \psi^{(k)} \)), where terms from lower-order perturbations \( \psi^{(k-1)},  \psi^{(k-2)}, \ldots, \psi^{(1)} \), along with their temporal and spatial derivatives, serve as an effective source for the perturbation at that order \( k \)\footnote{Here \( \psi = {}_{e} \psi + {}_{o} \psi \)}. From this point forward, we will use the operator \( \mathcal{L} \), the potential \( V \), and the perturbation scalars \( \chi^{(k)} \) to describe generic perturbations of order \( k \). Specifically, \( \psi^{(k)} \) will be reserved for gravitational perturbations, while \( \Phi^{(k)} \) will represent electromagnetic perturbations. Also, we will be focusing on the, first and second  perturbations, \( i = 1, 2 \)
\begin{equation}\label{generic perturbation scalar}
\begin{aligned}
    &\mathcal{L} := \partial_{xx} - \partial_{tt} - V(x), \\
    &\mathcal{L} \chi^{(i)}(t, x) = S^{(i)}(t, x),
\end{aligned}
\end{equation}
where \( S^{(i)}(t, x) \) is a generic source term at a perturbation order \( i \) and $x$ is the tortoise radial coordinates related to the areal radial coordinates through $dr=f(r) dx$, while $t$ is the time measured by a clock asymptotically far from the Schwarzschild BH.

\subsubsection{Laplace Transformation}
A common approach to solving the perturbation PDE is to apply the Laplace transform \cite{QNM_Nollert}  to its temporal dependence, resulting in an effective one-dimensional radial differential equation. The Laplace transform is defined as:
\begin{equation}
\begin{aligned}
L[f(t)] &= \tilde{f}(i\omega) = \int_0^\infty dt \, e^{-i\omega t} f(t), \\
L^{-1}[\tilde{f}(i\omega)] &= f(t) = \int_{i\epsilon -\infty}^{i\epsilon +\infty} d\omega \, e^{i\omega t} \tilde{f}(i\omega),
\end{aligned}
\end{equation}
where the integral is performed in the complex \(\omega\)-plane with \(\epsilon \to 0^+\). The perturbation operator in (\ref{generic perturbation scalar}) transforms as 
\[
\tilde{\mathcal{L}} = \partial_{xx} + \omega^2 - V(x),
\]
where the initial conditions of the system effectively acts as an additional source term for resulting ODE. To fully solve the problem up to second order in perturbation, two initial conditions must be specified for both the first- and second-order solutions
\begin{equation}
\begin{aligned}
\chi^{(i)}(x, 0) &= I^{(i)}(x), \\
\partial_t \chi^{(i)}(x, t) \big|_{t=0} &= K^{(i)}(x),
\end{aligned}
\end{equation}
where \( I^{(i)}(x) \) is the initial value, and \( K^{(i)}(x) \) is the initial rate of change for the \( i \)-th order perturbation.
\vspace{1mm}

\noindent In addition to the initial conditions, two boundary conditions in \( x \) are required. For astrophysically relevant scenarios, QNM boundary conditions are used. Specifically, we impose purely ingoing asymptotic (left BC) behavior near the horizon (\( x \to -\infty \)) and purely outgoing asymptotic (right BC) behavior at infinity (\( x \to \infty \)):
\begin{equation}
\begin{aligned}
&\tilde{\chi}^{(i)}(x, \omega) \to e^{i\omega x}, \quad x \to \infty, \\
&\tilde{\chi}^{(i)}(x, \omega) \to e^{-i\omega x}, \quad x \to -\infty,
\end{aligned}
\end{equation}
from now on we drop the tilde, so for instance \(\tilde{\chi}(x, \omega)\) becomes \(\chi(x, \omega)\). Then, the equation (\ref{generic perturbation scalar}) becomes
\begin{equation}\label{generic perturbation scalar laplace transformed}
\begin{aligned}
&\mathcal{L} \chi^{(i)}(\omega, x) = \mathcal{S}^{(i)}(\omega, x),\\
&\mathcal{S}^{(i)}(\omega, x) = S^{(i)}(\omega, x) - i \omega I^{(i)}(x) - K^{(i)}(x).
\end{aligned}
\end{equation}
\subsubsection{Green Function}
One effective approach to tackle a linear, one-dimensional, second-order inhomogeneous equation is by using the Green's function method, specifically in the context of BHPT\cite{szpak2004QNM,Andersson_1997}. The Green's function \( G(x, x', \omega) \) is defined by the equation
\begin{equation}
\mathcal{L} G(x, x', \omega) = \delta(x - x') .
\end{equation}
It can be expressed in terms of the solutions of the homogeneous equation \( \chi_{H} \)
\begin{equation}
G(x, x', \omega) = \frac{\chi_{H}^{+}(x_{>}, \omega) \chi_{H}^{-}(x_{<}, \omega)}{W(\omega)},
\end{equation}
where \( x_{>} = \max(x, x') \), \( x_{<} = \min(x, x') \), \( W(\omega) \) is the Wronskian of the solutions, and \( \chi_{H}^{+} \) (\( \chi_{H}^{-} \)) represent the homogeneous solution that satisfies the right (left) boundary condition. The solution to the inhomogeneous equation can then be written as:
\begin{equation}
\chi^{(i)}(x, \omega) = \int_{-\infty}^{\infty} dx' \, G(x, x', \omega) \mathcal{S}^{(i)}(x', \omega).
\end{equation}
\subsection{GEM Modes}\label{GEM Modes}
We now focus on gravitational perturbations. The overall effective source term for gravitational perturbation, \( {}_{G}\mathcal{S}^{(i)}(x, \omega) \), is given by:

\begin{equation}
\begin{aligned}
    {}_{G}\mathcal{S}^{(i)}(x, \omega) &= {}_{G} S^{(i)}_{EM}(x, \omega) + {}_{G} S^{(i)}_{Others}(x, \omega) \\
    &\quad - i \omega {}_{G} I^{(i)}(x) - {}_{G} K^{(i)}(x),
\end{aligned}
\end{equation}
where \( {}_{G} S^{(i)}_{EM}(x, \omega) \) represents the source term contribution from the electromagnetic field. Focusing on this contribution, we have:

\begin{equation}
{}_{EM}\psi^{(i)}(x, \omega) = \int_{-\infty}^{\infty} dx' \, {}_{G} G(x, x', \omega) {}_{G} S^{(i)}_{EM}(x', \omega).
\end{equation}
Here, interesting things happen. The term \( {}_{G} S^{(i)}_{EM}(x, \omega) \) is expected to depend on both \( \Phi(x, \omega)\) and \( \Phi(x, \omega) \times \Phi(x, \omega) \), so it should subsequently include terms linear and quadratic in the electromagnetic QNMs, similar to the case with the GG modes. Thus, it should be proportional to the electromagnetic QNM and their quadratic QNMs frequencies, which we refer to collectively as \( \Omega_{k} \),
\begin{equation}
{}_{G} S^{(i)}_{EM}(x, t) \propto e^{-i \Omega_{k} t},
\end{equation}
For the Laplace transform of this part, we get\footnote{Here we have made that assumption that Laplace transformation will only pick up poles due to the exponents.}
\begin{equation}
{}_{G} S^{(i)}_{EM}(x, \omega) \propto \frac{1}{\omega - \Omega_{k}}.
\end{equation}
Moreover, at higher orders, some source terms will be proportional to \( \Phi(x,\omega) \times \psi(x,\omega) \). In this sense, we can think of the new QNM spectrum of the Einstein-Maxwell system as an expansion/or an outer product of EM and G modes as well; not just linear and quadratic EM modes. 
\subsubsection{Inverse Laplace Transformation}
Back to our Green's function analysis, we can proceed with an inverse Laplace transformation (ILT) of the gravitational scalar\cite{QNMsILT} to obtain the time-domain expression \(\psi^{(i)}(x, t)\):
\begin{equation}
\begin{aligned}
    \psi^{(i)}(x, t) &= \int_{i\epsilon -\infty}^{i\epsilon +\infty} d\omega \, e^{i\omega t} \int_{-\infty}^{\infty} dx'\big[ {}_{G} S^{(i)}_{EM}(x', \omega)\\ &+ {}_{G} S^{(i)}_{GG}(x', \omega)- i \omega {}_{G} I^{(i)}(x') - {}_{G} K^{(i)}(x') \big]\\
    &\times \frac{\psi_{H}^{+}(x_{>}, \omega) \psi_{H}^{-}(x_{<}, \omega)}{W_{G}(\omega)}. \\
\end{aligned}
\end{equation}
From the singular structure of the ODE governing \(\chi\) (and consequently \(\psi\) and \(\Phi\)), \(\chi(x, \omega)\) is expected to be an analytic function over the domain \(x \in \mathbb{R}\) regardless of \(\omega\) \cite{ronveaux_2007}. Thus, we will proceed assuming that \(\psi_{H}^{+}(x, \omega)\) (and \(\psi_{H}^{-}(x, \omega)\)) would not contribute to the poles of the integral in the complex frequency domain. Hence, the only contribution will come from \({}_{G} S^{(i)}_{EM}(x', \omega)\), \( {}_{G} S^{(i)}_{GG}(x', \omega)\), and the zeros of \(W_G(\omega)\). Given the linearity of the ILT, we can focus on the pole of \(S^{(i)}_{EM}(x', \omega)/W_G(\omega)\) without any loss of generality. Accordingly, we can distinguish the G  and GEM modes, for instance, by focusing on second order gravitational perturbation
\begin{equation}
\begin{aligned}
    \psi^{(2)}(x, t) &= {}_{GG}\psi^{(2)}(x, t) + {}_{EM}\psi^{(2)}(x, t), \\
\end{aligned}
\end{equation}
where
\begin{equation}\label{psi EM contribuation}
\begin{aligned}
&{}_{EM}\psi^{(2)}(x, t) = \sum_{k} \frac{e^{i \Omega_k t}}{W(\Omega_k)} \int_{-\infty}^{\infty} dx' \bar{S}_{EM}(x',\Omega_k) \\
&\quad \times \psi_{H}^{+}(x_>, \Omega_k) \psi_{H}^{-}(x_<, \Omega_k),
\end{aligned}
\end{equation}
where $\bar{S}_{EM}$ is the residual of $S^{(2)}_{EM}$ after applying the Residue theorem. Moreover, even if we switched off the initial conditions and GG source ${}_{G} S^{(2)}_{GG}$, the $S^{(2)}_{EM}$ will still induce corrections on the first order G modes as in the case with GG modes, 
\begin{equation}\label{psi gravity contribuation}
\begin{aligned}
&{}_{G}\psi^{(1)}(x, t) = \sum_{n} C_n e^{i \omega_n t} \psi_{H}^{\pm}(x, \omega_n) \\
&C_n = \frac{1}{W'(\omega_n)} \int_{-\infty}^{\infty} dx' \left[ K^{(1)}(x') - i \omega_n I^{(1)}(x') \right. \\
&\quad \left. + S^{(1)}_{EM}(x', \omega_n) \right] \psi_{H}^{\pm}(x', \omega_n),
\end{aligned}
\end{equation}
where $\psi_{H}^{\pm}$ satisfy both Right and Left BC simultaneously. 
\vspace{1mm}

\noindent Generically, since we are dealing with a retarded Green's function, the problem respects a causal structure. In this sense, for a signal observed at an event $(t,x)$, there is an effective domain of support from the source in the $(t', x')$ plane, which is a subset of the latter space. This causal structure, as discussed in \cite{szpak2004QNM,Hui_2019}, ensured from treating the problem properly as an initial value problem using LT, thereby ensuring the regularity of the perturbations. Although, in some cases — such as asymptotic solutions or with simple potentials like the Dirac delta function — the causal part of the wavefunction factors out, and it is generally encoded in the Green's function itself. For more on the structure of Green's functions and the properties of QNMs, the reader is referred to \cite{Lagos_2023_GreenFunctionAnalysis_Quadratic_Diracdelta,Leaver_On_QNMs_Schwarzschild,Kokkotas_1999,ReviewArticle_QNM_Berti_Cardoso,Perrone_2024,Okuzumi_2008}.

\section{Toy Model}\label{toy_model}
It is well-known that the homogeneous solutions to the Regge-Wheeler, Zerilli, Press, and Teukolsky equations can be expressed in terms of Confluent Heun functions $H_C$ \cite{ ronveaux_2007} in the Laplace space \cite{Aly_2023,borissov2010exact,Fiziev}. Specifically, the solution to the Regge-Wheeler equation that satisfies the right boundary conditions is given by:
\begin{equation}
\begin{aligned}
 {}_{s}\chi_{H}^{+}(r, \omega)&= 2^{-1 - s} \, e^{-i \tilde{\omega} x} \left(\frac{r}{M}\right)^{1 + s}\\
&\times \, H_C\left(\scriptstyle{\lambda - (1 + s)(s - \tilde{\omega}), (1 + s) \tilde{\omega}, 1 - \tilde{\omega}, 1 + 2s, \tilde{\omega}, 1 - \frac{r}{2M}}\right) .
\end{aligned}
\end{equation}
In particular, the solutions to the electromagnetic perturbations are given by:
\begin{equation}
\begin{aligned}
 {}_{1}\chi_{H}^{+}(r, \omega)&= e^{-i \tilde{\omega} x} \left(\frac{r}{2M}\right)^2\\
&\times \, H_C\left(\scriptsize{-2 + \lambda + 2 \tilde{\omega}, 2 \tilde{\omega}, 1 - \tilde{\omega}, 3, \tilde{\omega}, 1 - \frac{r}{2M}}\right)\\
&+e^{-i \tilde{\omega} x} \, H_C\left(\lambda, 0, 1 - \tilde{\omega}, -1, \tilde{\omega}, 1 - \frac{r}{2M}\right)
 \end{aligned}
\end{equation}
where \(\tilde{\omega} = 4iM\omega\). However, applying boundary conditions at infinity in asymptotically flat spacetimes is more challenging compared to asymptotically de Sitter spacetimes, as discussed in \cite{Hatsuda_2020}. This difficulty arises because the Confluent Heun equation exhibits an irregular singularity of rank two at infinity. On the other hand, other analytical techniques such as MST \cite{Mano_1996, Shuhei_1997,Mano_1996RW}, semi-analytical ones such as WKB approximation \cite{WKB,WKB_Classical,GreenFunction_UsingLaplaceT_WKB,Lagos_2023_GreenFunctionAnalysis_Quadratic_Diracdelta} or even potential modifications like Pöschl-Teller potential \cite{ReviewArticle_QNM_Berti_Cardoso} are often employed to approximate solutions. In many instances, purely numerical methods are also used.
\vspace{1mm}

\noindent For illustration purposes, we will replace the Regge-Wheeler $V_{RW}$ and Zerilli $V_{Z}$ potential with a the Dirac delta potentials, \( V_G \delta(x - x_G) \) and \( V_{EM} \delta(x - x_{EM}) \) for gravitational and electromagnetic perturbation respectively, to examine the GEM modes more closely. The simplified model features a single G mode, EM mode, GG mode. This approximation serves as a useful tool for building intuition about the GEM mode mixing while being manageable analytically. In this section, we will rely heavily on our companion paper \cite{Fawzi_EM_idealdipole}.
\vspace{1mm}

\subsection{Dirac Delta Function Potential}\label{Dirac Delta Potential}
Away from the light ring located at \( r = 3M \), the potential diminishes, solutions asymptotically approach outgoing (ingoing) near the asymptotically flat region (horizon). This behavior, which is approximately true for exact potential at large distances, holds analytically for superlocalized potentials such as the Dirac delta function\cite{Lagos_2023_GreenFunctionAnalysis_Quadratic_Diracdelta}. The time-domain Green's function\footnote{The reader can find a detailed discussion about the properties of the flat Green's function component $G_F$, the  QNM component $G_{QNM}$, and the branch cut component $G_{B}$ in \cite{Leaver_On_QNMs_Schwarzschild,Lagos_2023_GreenFunctionAnalysis_Quadratic_Diracdelta,Kokkotas_1999}.} can be expressed as follows:
\begin{equation}
\begin{gathered}
G(t-t', r, r') = G_F(t-t', r, r') + G_Q(t-t', r, r'),\\
\end{gathered}
\end{equation}
The  Green’s function is expressed as
\begin{small}
\begin{equation}
 G(T, r, r') = -\frac{1}{2} \left[ \Theta(A) - \Theta(B) \right] 
- \frac{1}{2} e^{-\frac{V_{Mode}}{2}(B)} \Theta(B),
\end{equation}
\end{small}
where \( \Theta \) is the Heaviside step function. The derivative of the Green’s function with respect to \( r' \) is
\begin{small}
\begin{equation}
\begin{aligned}
f(r') &\frac{\partial \, G(t-t', r, r'')}{\partial r''} \bigg|_{r''=r'} 
= \frac{1}{2} \left[ s \delta(A) - \tilde{s} \delta(B) \right] \\
&\quad - \frac{\tilde{s}}{2} e^{-\frac{V_{Mode}}{2}(B)} 
\left[ \frac{V_{Mode}}{2} \Theta(B) - \delta(B) \right],
\end{aligned}
\end{equation}
\end{small}
where \( s(x - x') \) and \( \tilde{s}(x - x_{Mode}) \) are sign functions that take values of \( + \) for positive arguments and \( - \) for negative arguments. The parameters $A$ and $B$ encode the causal information, and are defined as
\begin{equation}
\begin{aligned}
A &= t - t' - |\bar{x} - \bar{x}'|, \\
B &= t - t' - |\bar{x}| - |\bar{x}'|,
\end{aligned}
\end{equation}
where $\bar{x} = x - x_{\text{Mode}}$, and $\text{Mode} = \{EM, G\}$. We can set $x_{EM}=0$ for simplicity without any loss of generality. $x_{G}$ needs to be slightly greater than $x_{EM}$, which is the case for exact potential. $x_{EM} \rightarrow x_{G}$ in the eikonal limit $l \xrightarrow{}\infty$.
\vspace{1mm}

\noindent The next step is to solve for both $\Phi^{(1)}$ and $\psi^{(i)}$. To proceed, we need to specify the nature of the current source for the electromagnetic part of our problem, as well as impose initial conditions on each perturbation scalar. For simplicity, we will assume homogeneous initial conditions for $\Phi^{(1)}$, and choose a radial ideal dipole  freely falling radially towards the black hole as the source of EM perturbation. 

\subsection{EM Mode: Ideal Dipole}\label{EM Mode Ideal Dipole}
In \cite{Fawzi_EM_idealdipole}, the radial ideal dipole is modeled as two charges, \( q \) and \( -q \), with a fixed separation \( \eta \) along their radial worldline into the black hole, each starting from rest at large distance $r_0$ and $r_0 +\eta $ respectively with \( E = 1 \) at angles $(\theta_{0},\phi_{0})$. The dipole moment is defined as $p=q \eta$. Electromagnetic interactions and self-interactions between the two particles along their worldlines are neglected. The resulting first order EM perturbation 
\begin{equation}
\begin{gathered}
{}_{e}\Phi^{dip(1)}_{lm}(r,t) = -\frac{q}{\lambda} Y_{lm}^* \eta \, \frac{\hat{\Phi}^{dip(1)}(r,t)}{2}, \\
\hat{\Phi}^{dip(1)} =  \hat{\Phi}^{dip(1)}_{Flat}(t,x) +A(t,x) e^{-\frac{V_{EM}}{2}(t - |x|)} \Theta(t-|x|).
\end{gathered}
\end{equation}
where $\hat{\Phi}^{dip(1)}_{Flat}(t,x)$ represents the flat part of the mode, while $A(t,x)$ is given by integrals $I_{Q1B}$ and $I_{Q2B}$, defined in \cite{Fawzi_EM_idealdipole}, capture how the 4-current source convolute with the QNM part of the green's function. Since, we are interested in how the EM perturbation contributions to the QNMs of the gravitational at first level of mixing--the order at which electromagnetic perturbation become relevant to gravitational one, we focus on $(t,x)$ at which the dipole wordline is interacting the potential peak at $x=0$.
\vspace{1mm}

\noindent Part of the flat contribution as well as the entire QNM contribution for a single point charge $q$ are provided by 
\begin{equation}
\begin{aligned}
A(t,x)&= \int_{t - |x|}^{t''} d u' \, (\beta - 1) f \left(\frac{3}{4} \frac{d f}{d r_p} + \frac{\tilde{s}}{4}  V_{EM} \right) e^{u' \frac{V_{EM}}{2}} \\
&- \int^{t''}_{x(r_0)} d v' \, (\beta + 1) f \left(\frac{3}{4} \frac{d f}{d r_p} + \frac{\tilde{s}}{4}  V_{EM} \right) e^{v' \frac{V_{EM}}{2}},
\end{aligned}   
\end{equation}
where $r_p$ is re-parameterized using $u'$($v'$) in each integral respectively, while $t''$ represents the time at which the particle passes the peak of the potential at $x=0$. It is worth noting that the contribution to $A(t,x)$ coming from the wordline near horizon $r=2M$ is negligible as $f\rightarrow 0$.
\vspace{1mm}

\begin{figure}
\includegraphics[width=8.5cm]{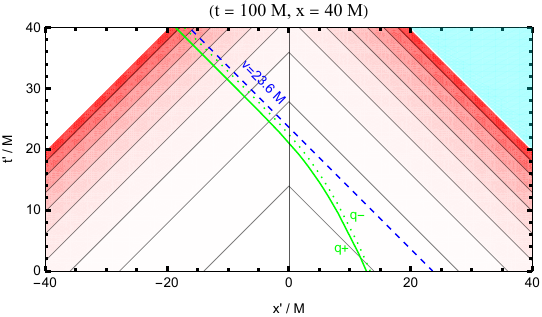}
\caption{This figure illustrates the QNM region \( B \) in gradient Red and the flat region \( B-A \) in light Blue, where the source consists of an ideal dipole with charges \( q \) and \( -q \) represented by the green solid and dotted curves, respectively. These curves asymptotically approach an ingoing null geodesic with \( v = 23.6 M \), shown as the blue dashed line with a slope of -45 degrees. The support region is chosen for a event $(t,x)$ with \( t = 100 M \)--for late-time scenario-- while \( x \) lies in the asymptotically flat region at \( x = 40 M \). The dipole's worldlines follow a characteristic behavior as they near the potential peak, transitioning toward null geodesics for late times. The geometry and dynamics depicted here are essential for understanding the electromagnetic QNMs generated by the dipole.}
\label{asymotitcally Null}
\end{figure}

\noindent Evaluating the integral in the above expression is challenging, as it requires inverting the relation \( u'(r_p) \)\footnote{The same holds for \( v'(r_p) \).}, which is difficult to accomplish analytically. Before addressing the dipole case, we first describe the validity of the approximation we are about to use. Indeed, we can expand \( r_p \) in terms of \( u' \) around \( u' = 0 \). This differs from the expansion used in \cite{Fawzi_EM_idealdipole}, yet we can leverage the fact that the geodesics of point particles asymptotically approach ingoing null geodesics near the potential peak, provided the particle started far enough away as shown in Fig. \ref{asymotitcally Null}.
\vspace{10mm}

\noindent Nevertheless, since only the left triangle of \( B \) contributes to how \( A(t, x) \) evolves as a function of both \( t \) and \( x \), we expand the logarithmic relation between \( x' \) and \( r_p \) as a polynomial in \( r_p \) around \( x'=0 \). Additionally, since \( v' \) remains constant—given that the geodesic is asymptotically ingoing null-like—we can express \( u' \) as an asymptotic polynomial in terms of \( r_p \). In this work, we only consider the linear contribution of this expansion. 
\vspace{2mm}

\noindent For one reason, in early times \( t \) such that the area of the support region \( (t-|x|)^2 \) is small, fewer terms in the expansion are needed. For instance, if \( |u| \) is small, then at early times of observing the signal when \( t \approx |x| \), the lowest-order expansion will suffice as region $B$ is more localized around $x=0$
\vspace{1mm}

\noindent On the other hand, this approximation should hold even at later times \( t \), since the logarithmic behavior of \( r \) is only significant near the horizon. But as we previously mentioned, the definition of \( A(t, x) \) remains unaffected by the particle's worldline there, making the approximation valid. Alternatively, the integration can be implemented fully numerically as in \cite{Fawzi_EM_idealdipole}.
\vspace{1mm}

\noindent Moving to the dipole case, since our primary interest lies in the electromagnetic QNM part of the dipole, varying this expression with respect to $r_0$, $\delta A(r_0)$, should capture this as well as part of the flat contribution \cite{Fawzi_EM_idealdipole}:
\begin{widetext}
\begin{equation}
\begin{aligned}
\delta A(r_{0}) = &\eta \left\{ c(r_0) +\int_{t - |x|}^{t''} du' \, e^{u' \frac{V_{EM}}{2}} \left[((\beta f)' - f') \left(\frac{3}{4} \frac{df}{dr_p} + \frac{\tilde{s}}{4}  V_{EM} \right) + (\beta - 1) f \left(\frac{3}{4} \frac{d^2 f}{dr_p^2} \right) \right] \right. \\
&\left.- \int_{x(r_{0})}^{t''} dv' \, e^{v' \frac{V_{EM}}{2}} \left[((\beta f)' + f') \left(\frac{3}{4} \frac{df}{dr_p} + \frac{\tilde{s}}{4}  V_{EM} \right) + (\beta + 1) f \left(\frac{3}{4} \frac{d^2 f}{dr_p^2} \right) \right] \right\}.
\end{aligned}
\end{equation}
\end{widetext}
Where \( c(r_0) \) represents the result of varying the integration limits\footnote{\(\eta\) is approximated to be constant along the dipole worldline.}. For a more detailed explanation of the flat part or the full derivation, we refer the reader to \cite{Fawzi_EM_idealdipole}.
\vspace{1mm}

\noindent We study electromagnetic perturbations which source gravitational ones, so we focus on the electromagnetic field profile in the region around $x=0$. This implies that we cannot simply truncate the expansion at the first few orders for late times. However, for illustrative purpose and to capture the gravitational perturbation at early time, we maintain at least a linear term in $u'$ in the $r_p$ expansion and integrate. Subsequently, $\delta A$ will have two contributions from the QNM support region defined by $B=0$: a flat part  and a QNM part. Their expression for $x>0$ is given by
\begin{widetext}
\begin{equation}
\begin{aligned}
e^{-\frac{V_{EM}}{2}u}\delta A \Bigg|_{Flat}=&\frac{a_{1\,+}}{(\alpha_{+}+ u)^{\frac{1}{2}}} + \frac{a_{2\,+}}{(\alpha_{+}+ u)} + \frac{a_{3\,+}}{(\alpha_{+}+ u)^{\frac{3}{2}}} + \frac{a_{4\,+}}{(\alpha_{+}+ u)^2}  +  \frac{a_{5\,+}}{(\alpha_{+}+ u)^{\frac{5}{2}}} + \frac{a_{6\,+}}{(\alpha_{+}+ u)^3}+ \frac{a_{7\,+}}{(\alpha_{+}+ u)^{\frac{7}{2}}}.
\end{aligned}
\end{equation}

\begin{equation}\label{QNM_Variation}
\begin{aligned}
e^{-\frac{V_{EM}}{2}u}\delta A \Bigg|_{QNM} =
& e^{-\frac{V_{\text{EM}}}{2} u} \Bigg( 
 b_0 + b_{+} \, \text{Ei}\left[ \frac{V_{\text{EM}} J_{+}}{2 H_{+}} + \frac{V_{\text{EM}} u}{2} \right]+ b_{-} \, \text{Ei}\left[ \frac{V_{\text{EM}} J_{-}}{2 H_{-}} + \frac{V_{\text{EM}} u}{2} \right] \Bigg)
\end{aligned}
\end{equation}
\end{widetext}
where $a_{i\, +}$, $\alpha_{+}$, $b_{0}$, $b_{+}$, and $d_{+}$, as well as $J_{+}$ and $H_{+}$, are constant of expansion and they parameterized by $M$ and $r_0$. The function $\text{Ei}$ represents the exponential integral function, while $\gamma$ denotes the incomplete gamma function. In that sense the amplitude of QNMs is the coefficient of the linear EM exponent in Eq. (\ref{QNM_Variation}). For the rest of the flat part of the EM perturbation, we refer the reader to \cite{Fawzi_EM_idealdipole}. Moreover, to obtain the variations for $x<0$, we simply replace $u \rightarrow v$.
\vspace{1mm}

\noindent Moving forward to the Faraday tensor $F_{\mu \nu}$, there is no monopole contribution \( F_{\text{monopole}} \) to \( F \), and the contribution due to the localized dipole, \( {}_{e2} F_{\text{int}} \), vanishes away from the dipole. Only the radiative part, \( {}_{e1} F_{\text{rad}} \), remains non-zero and is described by the even electromagnetic scalar perturbation \( \hat{\Phi}^{\text{dip}(1)}(r, t) \), which is given as follows
\begin{widetext}
\begin{equation}\label{dipole_Stress_Energy_Tensor}
\begin{aligned}
{}_{e1} F_{\text{rad}} &= -p \begin{pmatrix}
0 & ** & ** & ** \\
\left[\delta(\cos \theta -\cos \theta_0)\delta(\phi-\phi_0) - \frac{1}{4\pi}\right] \frac{\hat{\Phi}^{dip(1)}(t,r)}{r^2} & 0 & ** & ** \\
-\frac{\partial G_{S^2}}{\partial \theta} f \frac{\partial \hat{\Phi}^{dip(1)}(t,r)}{\partial r} & -\frac{\partial G_{S^2}}{\partial \theta} \frac{1}{f} \frac{\partial \hat{\Phi}^{dip(1)}(t,r)}{\partial t} & 0 & 0 \\
-\frac{\partial G_{S^2}}{\partial \phi} f \frac{\partial \hat{\Phi}^{dip(1)}(t,r)}{\partial r} & -\frac{\partial G_{S^2}}{\partial \phi} \frac{1}{f} \frac{\partial \hat{\Phi}^{dip(1)}(t,r)}{\partial t} & 0 & 0
\end{pmatrix}
\end{aligned}
\end{equation}
\end{widetext}
where $G_{S^{2}}$ is the angular Green's function for the Laplace operator defined on \( S^{2} \) sphere, and its expression can be found in \cite{Fawzi_EM_idealdipole}.

\subsection{GEM Mode: Ideal Dipole}\label{Gravtional IDeal dipole}
Finally, we are in a position to examine the behavior of the GEM mode. First, the sources for the Regge-Wheeler and Zerilli equations need to be calculated from the EM stress-energy tensor. This can be achieved by following a similar angular decomposition as in \cite{KarlMartel_2005_RW,Barack_2005} to calculate the electromagnetic contribution to these source terms ${}_J S_{o,e \, lm}$. The source terms should contain both linear EM modes, $\Omega_{1} = -i \frac{V_{\text{EM}}}{2}$, and quadratic modes, $\Omega_{2} = -i V_{\text{EM}}$, similar to the gravitational effective source terms in GG modes. From (\ref{dipole_Stress_Energy_Tensor}), the even and odd sources due to an ideal dipole moment $p$, freely falling towards the black hole along a radial trajectory, are given as follows
\begin{widetext}
\begin{equation}\label{sourceEVEN}
\begin{aligned}
&{}_{J} S_{o\, l'm'}= 0\\
&{}_{J} S_{e\, l'm'}= \Phi^{\text{dip}(1)} \frac{\partial \Phi^{\text{dip}(1)}}{\partial r}  \left[  \frac{1}{2 + \mu' } C_{l'm'}  - \frac{2 \,  \pi}{ \lambda'}  D_{l'm'} \right]  \frac{16 f}{ \Lambda'} + \Phi^{\text{dip}(1)\, 2} \,\, \frac{  4 \Lambda' g(r)  - 8 f (6 M + \Lambda' r) }{(\mu' + 2) \Lambda'^2 r^2}  f C_{l'm'} \,  \\
&+64 \pi^2 \left[ \left( \frac{\partial \Phi^{\text{dip}(1)}}{\partial t} \right)^2  \left( \frac{12 \,  M}{(\mu' + 2) \Lambda'^2} B_{l'm'} + \frac{l' \, }{\lambda \mu' r f} R_{l'm'} \right) 
+ \left( \frac{\partial \Phi^{\text{dip}(1)}}{\partial x} \right)^2 \left( \frac{12\, M}{(\mu' + 2) \Lambda'^2}  B_{l'm'} - \frac{l'\,}{\lambda \mu' rf} R_{l'm'} \right) \right]
\end{aligned}
\end{equation}
\end{widetext}
where $\mu = \lambda - 2$ and $\Lambda = \mu + \frac{6M}{r}$. The angular coefficients are reported in appendix \ref{Source for Gravitational toymodel} using the tensor harmonics defined in \ref{Tensor Harmonics}. The function $g(r)$ is defined as 
\[
g(r) = \mu' (2 - \mu') r + 12M (3 - \mu') - 84 \frac{M^2}{r} .
\]
\noindent We now focus on solving the second-order gravitational perturbation, again assuming a homogeneous initial condition on $\psi_{lm}^{(2)}$. In this case, the odd scalar doesn't receive any contribution from the electromagnetic field, ${}_{\text{EM}} \psi_{o \, lm}=0$, whereas the even scalar is influenced not only by the gravitational mode mixing but also by contributions from the electromagnetic field given by
\begin{widetext}
\begin{equation}
{}_{GG,EM}\psi^{(2)}_{e\,lm}\left(r^{\prime}, t^{\prime}\right)=\int_{-\infty}^{\infty} d t \int_{2 M}^{\infty} d r \, G\left(t- t^{\prime}, r, r^{\prime}\right) {}_{GG,J} \mathcal{S}_{e \, lm} .
\end{equation}
\end{widetext}
The full even solution is a superposition given by 
\[
\psi_{e \, lm}^{(2)} = {}_{GG} \, \psi_{e \, lm}^{(2)} + {}_{EM} \, \psi_{e \, lm}^{(2)}.
\]
\noindent In what follows, we will sample from both the linear, $\Omega_{1}$, and quadratic, $\Omega_{2}$, contributions of the EM modes to the source. The simplest case, is a  ${}_{J} S_{e}(t,x>0) \propto e^{-V_{\text{em}} u}$, this will just induce gravitational perturbation quadratic in EM mode in addition to linear term in G mode with amplitude dependent on the details of electromagnetic as well as gravitational potentials. Next, we can consider a source sample similar to the source term that was investigated in
\cite{Lagos_2023_GreenFunctionAnalysis_Quadratic_Diracdelta}, 
\begin{equation}\label{Logas_Source}
{}_{J} S_{e}(t,x>0) \propto  \frac{e^{-V_{\text{em}} u}}{(1+\zeta |x|)^n},
\end{equation}
where $n=1,2,3$. In second term in (\ref{sourceEVEN}), after expanding the coefficient around $|x|=0$, we get
\[
\begin{aligned}
\frac{  4 \Lambda' g(r)  - 8 f (6 M + \Lambda' r) }{(\mu' + 2) \Lambda'^2 r^2} &=
\frac{z_{1 \, l'm'}}{r} + \frac{z_{2\, l'm'}}{r^2} + \frac{z_{3\, l'm'}}{r^3}\\
&+ \frac{z_{4\, l'm'}}{1+ z_{5\, l'm'} r} + \frac{z_{6\, l'm'}}{(1+ z_{7\, l'm'} r)^2}
\end{aligned}
\]
where $z_{i \, lm}$ are constants. We focus only on the $n=2$ case in order to include the GG mode from \cite{Lagos_2023_GreenFunctionAnalysis_Quadratic_Diracdelta}, with the simple transformation $V_{\text{EM}} \rightarrow V_G$ and $\zeta \rightarrow V_G$, when applied to (\ref{Logas_Source}). However, the cases for $n=1$ and $n=3$ are similar. We focus only on the QNM contribution, supported by the region $B$, as

\begin{equation}
\begin{aligned}
    \psi_{QNM}^{(2)} \propto &- \frac{8 e^{-\frac{2 V_{\text{EM}}}{\zeta}}}{(2 V_{\text{EM}} - V_{G}) \zeta^2} e^{-V_{\text{EM}} u} 
    \Bigg\{- 2 V_{\text{EM}} \, \text{Ei}\left[\frac{2 V_{\text{EM}}}{\zeta}\right] \\
    &+e^{\frac{2 V_{\text{EM}}}{\zeta}} \zeta  + 2 V_{\text{EM}} \, \text{Ei}\left[ u V_{\text{EM}} + \frac{2 V_{\text{EM}}}{\zeta} \right]
    \Bigg\} \\
    &+ \frac{8 e^{-\frac{V_{G}}{\zeta}}}{(2 V_{\text{EM}} - V_{G}) \zeta^2} e^{-\frac{V_{G}}{2} u} 
    \Bigg\{ 
        - V_{G} \, \text{Ei}\left[\frac{V_{G}}{\zeta}\right] \\
    &+ e^{\frac{V_{G}}{\zeta}} \zeta + V_{G} \, \text{Ei}\left[\frac{u V_{G}}{2} + \frac{V_{G}}{\zeta}\right] 
    \Bigg\},
\end{aligned}
\end{equation}

\noindent Although the linear EM, $\Omega_{1}$, mode doesn't appear in those samples, along with other source term samples, it demonstrates that the second-order gravitational perturbations QNM $\psi_{QNM}^{(2)}$ exhibit both a quadratic EM frequency, $\Omega_{2}$, and a linear G frequency, $\omega_{1}$, typically with a $(t, x)$-dependent amplitude. However, for $\frac{|u|}{V_{\text{EM}} V_G} \gg 0$, these terms might suggest a tail behavior, similar to what is shown in \cite{Lagos_2023_GreenFunctionAnalysis_Quadratic_Diracdelta}:
\begin{widetext}
\begin{equation}\label{tail}
\begin{aligned}
&-\frac{ 2 V_{\text{EM}} e^{-\frac{2 V_{\text{EM}}}{\zeta}}}{(2 V_{\text{EM}} - V_G) \zeta^2} e^{-V_{\text{EM}} u}   \, \text{Ei}\left( u V_{\text{EM}} + \frac{2 V_{\text{EM}}}{\zeta} \right) 
+ \frac{ V_G e^{-\frac{V_G}{\zeta}}}{(2 V_{\text{EM}} - V_G) \zeta^2} e^{-\frac{V_G u}{2}}  \, \text{Ei}\left( \frac{u V_G}{2} + \frac{V_G}{\zeta} \right)\\
&\simeq\frac{1}{{(2 V_{\text{EM}} - V_G) \zeta^2}} \left\{
-\frac{2 V_{\text{EM}}}{\left( u V_{\text{EM}} + \frac{2 V_{\text{EM}}}{\zeta} \right)^2}
- \frac{2 V_{\text{EM}}}{\left( u V_{\text{EM}} + \frac{2 V_{\text{EM}}}{\zeta} \right)^3} 
+ \frac{4 V_G}{\left( u V_G + \frac{2 V_G}{\zeta} \right)^2} 
+ \frac{16 V_G}{\left( u V_G + \frac{2 V_G}{\zeta} \right)^3}
\right\}\\
&\approx \frac{2}{u^2} \frac{1}{V_{E M} V_G \xi^2}+\frac{4}{u^3} \frac{2 V_{E M}+V_{G}}{V_{E M}^2 V_G^2 \xi^2}\\
\end{aligned}
\end{equation}
\end{widetext}
Although the approximation for the dipole's electromagnetic perturbation used in subsection \ref{EM Mode Ideal Dipole} linearizes \( u \) in \( r_p \), the geodesic of the dipole asymptotically becomes null-like near the horizon, as discussed earlier. Thus, we might anticipate that this approximation is not too inaccurate for late \( t \) as well since only the left triangle, of the $B$ region in $(x,t)$ plane, matters in that case. Still, to safely confirm that we need to consider a higher order expansion in $u$ or a better way of implementing the integration.
\vspace{1mm}

\noindent Next, we consider a sample that is linear only in the EM mode, $\Omega_{1}$, for example:
\[
{}_{J} S_{e}(t,x>0) \propto b_{0} e^{-\frac{V_{\text{em}}}{2} u} \left[ \frac{a_{1\,+}}{(\alpha_{+}+u)^{\frac{1}{2}}}\right] .
\]
There will be two contributions: one to the flat $\psi_{eQNM(0)}^{(2)}$ part and one to the linear G mode $\psi_{eQNM(1)}^{(2)}$ 
\begin{widetext}
\begin{equation}\label{eQNM0}
\begin{aligned}
{}_{EM} \, \psi_{eQNM(1)}^{(2)} \propto \frac{b_{0} \sqrt{2 \pi}e^{-\frac{V_{G}}{2} u}}{V_{G}^{4} \sqrt{V_{EM}-V_{G}}}  \Bigg[ 
& a_{1+} e^{\frac{\alpha_{+}}{2} (V_{\text{EM}} - V_{G})} 
\left\{ \text{Erf}\left[\sqrt{\frac{V_{\text{EM}} - V_{G}}{2}} \sqrt{\alpha_{+}}\right] 
- \text{Erf}\left[\sqrt{\frac{V_{\text{EM}} - V_{G}}{2}} \sqrt{u + \alpha_{+}}\right] \right\}
\Bigg]
\end{aligned}
\end{equation}

\begin{equation}\label{eQNM1}
\begin{aligned}
\psi_{eQNM(0)}^{(2)} \propto \frac{ b_{0} \sqrt{2 \pi}}{V_{G}^{4} \sqrt{V_{\text{EM}}}}& \Bigg[ 
& a_{1+} e^{\frac{V_{\text{EM}} \alpha_{+}}{2}} \left\{ 
\text{Erf}\left[\sqrt{\frac{V_{\text{EM}}}{2}} \sqrt{u + \alpha_{+}}\right] - \text{Erf}\left[\sqrt{\frac{\alpha_{+} V_{\text{EM}}}{2}}\right] \right\},
\Bigg]
\end{aligned}
\end{equation}
\end{widetext}
 in this sample, the amplitude of the linear G mode, which depends on $t - |x|$, exhibits corrections that are influenced by the EM perturbation. This behavior is similar to the analysis in \cite{Lagos_2023_GreenFunctionAnalysis_Quadratic_Diracdelta}. There will be regions of spacetime where the amplitude loses its sensitivity to $(t, x)$, yet the factor $\frac{1}{\sqrt{V_{\text{EM}} - V_G}}$ will continue to leave imprints due to the electromagnetic-gravitational coupling. The same conclusion applies to the flat part. To obtain the expression for $x < 0$, the flat part remains the same, but for the linear mode, the transformation $e^{-\frac{V_{\text{EM}}}{2}u} \rightarrow e^{-V_{\text{EM}} x}$ is applied.
Next, we can consider a source sample that is quadratic in the EM modes, $\Omega_{2}$. This will give rise to a quadratic term in EM, ${}_{\text{EM}} \, \psi_{eQNM(2)}^{(2)}$, a linear term in G, ${}_{\text{EM}} \, \psi_{eQNM(1)}^{(2)}$, and a flat term, ${}_{\text{EM}} \, \psi_{eQNM(1)}^{(2)}$, as follows:
\begin{widetext}
\[
{}_{J} S_{e}(t,x>0) \propto b_{0} e^{-uV_{\text{EM}} } \left( b_{+} \, \text{Ei}\left[\tilde{\alpha}_{+} + \frac{V_{\text{EM}} u}{2}\right] 
\right)\]
\begin{equation}
\begin{aligned}
{}_{EM} \, \psi_{eQNM(2)}^{(2)} \propto \frac{4 b_{0} e^{-u V_{\text{EM}}}}{V_{\text{EM}}(2 V_{\text{EM}} - V_{G})} \left\{b_{+}  \text{Ei}\left[\frac{1}{2} V_{\text{EM}}(u+\tilde{\alpha}_{+})\right] \right\} 
\end{aligned}
\end{equation}
\begin{equation}
\begin{aligned}
{}_{EM} \, \psi_{eQNM(1)}^{(2)} \propto \frac{-8 b_{0} e^{-\frac{u V_{G}}{2}}}{(2 V_{\text{EM}} - V_{G}) V_{G}} \Bigg[ 
&b_{+}e^{\tilde{\alpha}_{+}\left(V_{\text{EM}}- \frac{V_{G}}{2} \right)} \left\{\text{Ei}\left[-\frac{1}{2}(V_{\text{EM}} - V_{G})(u+\tilde{\alpha}_{+})\right] 
- \text{Ei}\left[-\frac{1}{2}(V_{\text{EM}} - V_{G}) \tilde{\alpha}_{+}\right]\right\}\\&+b_{+}\text{Ei}\left[\frac{V_{\text{EM}} \tilde{\alpha}_{+}}{2}\right] \Bigg]
\end{aligned}
\end{equation}
\begin{equation}
\begin{aligned}
{}_{EM} \, \psi_{eQNM(0)}^{(2)} \propto  \frac{4 b_{0}}{V_{\text{EM}} V_{G}} \Bigg[& b_{+} \text{Ei}\left[\frac{V_{\text{EM}} \tilde{\alpha}_{+}}{2}\right]-b_{+} e^{V_{\text{EM}} \tilde{\alpha}_{+}} \left\{ \text{Ei}\left[-\frac{V_{\text{EM}} \tilde{\alpha}_{+}}{2}\right] 
- \text{Ei}\left[-\frac{1}{2} V_{\text{EM}}(u+\tilde{\alpha}_{+})\right]\right\} \Bigg] \\
\end{aligned}
\end{equation}
\end{widetext}
At early times, the linear G and quadratic EM modes have an asymptotic constant amplitude due to the $Ei$ expansion for $u \rightarrow 0$. Based on the same expansion, the flat part should also initially remain constant. However, an interesting behavior of this part of the GEM mode appears at late times when $
\frac{|u|}{V_{\text{EM}} V_G} \gg 0$. All contributions, ${}_{\text{EM}} \, \psi_{eQNM(0)}^{(2)}$, ${}_{\text{EM}} \, \psi_{eQNM(1)}^{(2)}$, and ${}_{\text{EM}} \, \psi_{eQNM(2)}^{(2)}$, from this source sample switch to a linear EM mode that is suppressed with an additional polynomial decay, similar to the tail part in (\ref{tail}).
\vspace{1mm}

\noindent However, one difference that we did not account for in this calculation is that the peak of the gravitational potential should be slightly greater than its electromagnetic counterpart, $x_{G} > x_{\text{EM}}$. Hence, the support regions $B_{\text{EM}}$ and $B_{G}$ are intrinsically different. To evaluate the integral in the closed form, we ignored this difference and assumed $x_{G} = x_{\text{EM}}$, which is only true in the eikonal limit, as mentioned in \ref{Dirac Delta Potential}. The resultant causality function of ${}_{EM} \psi^{(2)}$ would then be $\Theta(t - |x|)\Theta(t - |x - x_{G}|)$.

\section{Conclusion and Future Outlook}\label{Conclusion}
The main purpose of this work was to expand and refine theoretical predictions for the GW signals by taking into account the effects of the strong electromagnetic fields during an inspiral, collapse or a merger phase of massive objects. A simple order of magnitude estimate suggests that the effects of gravitational and electromagnetic mode mixing could be of some importance for a merger of a NS with a SMBH.
\vspace{1mm}

\noindent Using BHPT, we investigated the effect of the electromagnetic coupling on the QNMs of the system. In section \ref{GEM Modes}, we showed that, within the minimally coupled Einstein-Maxwell theory, the gravitational QNMs spectrum will expand to include both the linear and quadratic electromagnetic QNMs. Moreover, at higher order in PT, we expect higher level of mixing including having modes quadratic in electromagnetic and gravitational modes similar to the ones discussed in \cite{Fawzi_EM_idealdipole} and mentioned in section \ref{Discussion} below.
\vspace{1mm}

\noindent We recast the electromagnetic stress-energy tensor to align with the Regge-Wheeler spherical decomposition by projecting it on the sphere following \cite{KarlMartel_2005_RW}. Subsequently, this simplify the identification of which pairs of the angular modes of electromagnetic $(l_{1},m_{1})$ and $(l_{2},m_{2})$ modes will source a gravitational mode $(l,m)$.
\vspace{1mm}

\noindent In section \ref{toy_model}, to illustrate and build intuition about the effects of electromagnetic-gravitational coupling, we replaced the actual potentials in the Regge-Wheeler and Zerilli equations, which describe the two degrees of freedom for each perturbation field, with a Dirac delta function potential. This situation is easier to handle analytically, since it results in only one gravitational, one electromagnetic, and one quadratic gravitational QNM. These simplifications allow to perform the calculations for the GEM modes semi-analytically.
\vspace{1mm}

\noindent We used the approximate formula for the electromagnetic perturbation caused by an ideal dipole radially falling towards a black hole in the Dirac delta function potential, reported in detail in \cite{Fawzi_EM_idealdipole} and summarized in subsection \ref{EM Mode Ideal Dipole} for completeness. We then used these perturbations to compute the induced GEM modes from various samples of the electromagnetic source in the gravitational perturbation equation. As anticipated, both linear and quadratic electromagnetic QNMs frequencies showed up in induced gravitational QNMs. Additionally, as discussed in section \ref{GEM Modes}, the linear gravitational modes exhibit amplitudes that generally depend on the details of both the electromagnetic and gravitational potentials. Furthermore, at the late stages of the signal, we found that the gravitational perturbation exhibits polynomial tails due to the electromagnetic perturbation, similar to the findings in \cite{Lagos_2023_GreenFunctionAnalysis_Quadratic_Diracdelta} for gravitational-gravitational coupling.
\vspace{1mm}

\vspace{1mm}

\noindent LISA will focus on the inspiral phase of GW signals, in contrast to LIGO, which concentrates more on the ringdown phase. Extending this investigation to search for EM frequencies embedded within the EMRI signal requires a slightly different approach. For example, considering a charged perturber of mass $m$ and charge $q$ circling around a supermassive black hole of mass $M$, we can analytically model the inspiral phase using BHPT. This allows us to identify regions in the parameter space $(q,m,M)$, where GG modes and GEM modes are of comparable magnitude,
\[
{}_{GG} \, \psi_{e \, lm}^{(2)} \sim {}_{EM} \, \psi_{e \, lm}^{(2)}.
\]
A comparison between \({}_{G} S_{EM}\) and \({}_{G} S_{GG}\) will be important in finding out whether this region overlaps with LISA’s observation window, indicating the relevance of such events. Consequently, when \( {}_{EM} \, \psi_{e \, lm}^{(2)} \) should not be ignored when modeling the inspiral signal. However, handling stellar-mass mergers presents more challenges, as the remnant can only be described using PT in the late stages. In such cases, a numerical approach may provide more insight, such as fitting waveforms from numerical simulations or using numerical guidance to model the initial conditions. We will attempt further investigation related to those directions in future work, building on the foundation laid in this paper.
\vspace{1mm}

\noindent Additionally, in \cite{Fawzi_EM_idealdipole}, the authors speculated that at higher orders of electromagnetic perturbations \( \Phi^{(j)}, j \geq 2 \), gravitational perturbations will also contribute to exciting higher-order electromagnetic modes. At second-order electromagnetic QNMs, for instance, stronger coupling between the two fields, akin to what occurs in the gravitational case but in higher orders can take place. In particular, if a linear gravitational QNM has frequency \( \omega^{(1)}_i \) and a linear electromagnetic QNM has frequency \( \Omega^{(1)}_j \), the second-order electromagnetic spectrum will feature mixed gravitational-electromagnetic modes with frequencies \( \Omega^{(2)}_i = \Omega^{(1)}_j \pm \omega^{(1)}_k \). However, purely quadratic electromagnetic modes will not be present at that order in contrast with gravitational case at its simplest level of coupling with electromagnetic QNMs.
\vspace{1mm}

\noindent Since gravitational effects are expected to dominate in astrophysical scenarios, this suggests that gravitational imprints in electromagnetic signals should be more pronounced compared to the other way around. Yet, both phenomena can serve the same purpose: testing minimal coupling in GR or more complex couplings in modified gravity theories. Plausibly, if both effects are significant, it could offer a key opportunity for multi-messenger astronomy.
\vspace{1mm}

\noindent

\begin{acknowledgments}
We acknowledge the partial support of Dr. D.S. through the US National Science Foundation under the Grant No.  PHY-2310363. Additionally, we express our gratitude to Macarena Lagos, Ariadna Ribes Metidieri, Ahmed Elshahawy, A.K. Gorbatsievich, and Stanislav Komarov for their insightful discussions. Special thanks to Jacob Fields and Ish Gupta for their invaluable guidance and comments on this manuscript.
\end{acknowledgments}
\appendix
\begin{widetext}
\section{4D Tensor Harmonics}\label{Tensor Harmonics}
The tensor harmonics $\mathcal{Y}_{\mu\nu(7)}$,$\mathcal{Y}_{\mu\nu(8)}$,and $\mathcal{Y}_{\mu\nu(9)}$ are odd while the rest are even.

\begin{equation}
\begin{array}{ccc}
\mathcal{Y}_{\mu\nu(1)} = \begin{pmatrix}
1 & 0 & 0 & 0 \\
0 & 0 & 0 & 0 \\
0 & 0 & 0 & 0 \\
0 & 0 & 0 & 0
\end{pmatrix} Y_{lm}, &
\mathcal{Y}_{\mu\nu(2)} = \begin{pmatrix}
0 & 1 & 0 & 0 \\
1 & 0 & 0 & 0 \\
0 & 0 & 0 & 0 \\
0 & 0 & 0 & 0
\end{pmatrix} Y_{lm}, &
\mathcal{Y}_{\mu\nu(3)} = \begin{pmatrix}
0 & 0 & 0 & 0 \\
0 & 1 & 0 & 0 \\
0 & 0 & 0 & 0 \\
0 & 0 & 0 & 0
\end{pmatrix} Y_{lm}, 
\end{array}
\end{equation}

\begin{equation}
\begin{array}{ccc}
\mathcal{Y}_{\mu\nu(4)} = \begin{pmatrix}
0 & 0 & 0 & 0 \\
0 & 0 & 0 & 0 \\
0 & 0 & 1 & 0 \\
0 & 0 & 0 & \sin^2\theta
\end{pmatrix} Y_{lm}, &
\mathcal{Y}_{\mu\nu(5)} = \begin{pmatrix}
0 & 0 & \dot{Y}_{lm} &  Y_{lm}^{\prime} \\
0 & 0 & 0 & 0 \\
** & 0 & 0 & 0 \\
** & 0 & 0 & 0
\end{pmatrix}, &
\mathcal{Y}_{\mu\nu(6)} = \begin{pmatrix}
0 & 0 & 0 & 0 \\
0 & 0 & \dot{Y}_{lm} & Y_{lm}^{\prime} \\
0 & ** & 0 & 0 \\
0 & ** & 0 & 0
\end{pmatrix},
\end{array}
\end{equation}

\begin{equation}
\begin{array}{ccc}
\mathcal{Y}_{\mu\nu(7)} = \begin{pmatrix}
0 & 0 & -\csc\theta \, Y_{lm}^{\prime} & \sin\theta \, \dot{Y}_{lm} \\
0 & 0 & 0 & 0 \\
** & 0 & 0 & 0 \\
** & 0 & 0 & 0
\end{pmatrix}, &
\mathcal{Y}_{\mu\nu(8)} = \begin{pmatrix}
0 & 0 & 0 & 0 \\
0 & 0 & -\csc\theta \, Y_{lm}^{\prime} & \sin\theta \, \dot{Y}_{lm} \\
0 & ** & 0 & 0 \\
0 & ** & 0 & 0
\end{pmatrix},
\end{array}
\end{equation}

\begin{equation}
\begin{array}{cc}
\mathcal{Y}_{\mu\nu(9)} = \begin{pmatrix}
0 & 0 & 0 & 0 \\
0 & 0 & 0 & 0 \\
0 & 0 & -\csc\theta \, D_1[Y_{lm}] & \sin\theta \, D_2[Y_{lm}] \\
0 & 0 & ** & \sin\theta \, D_1[Y_{lm}]
\end{pmatrix}, &
\mathcal{Y}_{\mu\nu(10)} = \begin{pmatrix}
0 & 0 & 0 & 0 \\
0 & 0 & 0 & 0 \\
0 & 0 & D_1[Y_{lm}] & D_2[Y_{lm}] \\
0 & 0 & ** & -\sin^2\theta \, D_1[Y_{lm}]
\end{pmatrix}
\end{array}
\end{equation}
where 
\begin{align}
D_1[f] &= \frac{\partial f}{\partial \phi \, \partial \theta} - \cot\theta \, \frac{\partial f}{\partial \phi}, \\
D_2[f] &= \frac{\partial^2 f}{\partial \theta^2} + \frac{\lambda}{2} f,
\end{align}
\noindent
The orthogonality of the vector harmonics \(\mathcal{Y}^{\mu}_{(i)}\) on a 2-sphere \( S^2 \) as follows:
\begin{equation}
\int_{S^2} \mathbf{\mathcal{Y}}^{\mu\nu}_{(i)lm} \mathbf{\mathcal{Y}}_{\mu\nu(j)l'm'} \, d\Omega = \delta_{ij} \delta_{ll'} \delta_{mm'} \mathcal{N}_{i}(r) ,
\end{equation}
where $\sqrt{\mathcal{N}_{i}(r)}$ is the norm of the $i$th tensor.

\section{Source for Gravitational Perturbation}\label{Source for Gravitational toymodel}
\begin{equation}
P^r = \frac{8 \pi r^2}{\lambda'} \int d\Omega \, T^{\mu \nu} \mathcal{Y}_{\mu \nu (8)}^* = -\Phi^{\text{dip}(1)} \frac{\partial \Phi^{\text{dip}(1)}}{\partial r} \frac{8 \pi f A_{l' m'}}{\lambda'}
\end{equation}

\begin{equation}
P^t = \frac{8 \pi r^2}{\lambda'} \int d\Omega  \, T^{\mu \nu} \mathcal{Y}_{\mu \nu (7)}^{*} = \Phi^{\text{dip}(1)} \frac{\partial \Phi^{\text{dip}(1)}}{\partial t} \frac{8 \pi A_{l' m'}}{\lambda' f}
\end{equation}

\begin{equation}
A_{l' m'} = -4 \pi \csc\theta_0 \sum_{l m} \frac{Y_{l m}^*}{\lambda'} \left\{ Y_m' \dot{Y}_{l' m'}^* - \dot{Y}_{l m} Y_{l' m'}'^* \right\}
\end{equation}

\begin{equation}
Q^{tt} = 8 \pi \int d\Omega \, T^{\mu \nu} \mathcal{Y}_{\mu \nu (1)}^* = \frac{\Phi^{\text{dip}(1)\, 2}}{r^2} C_{l'm'} + 16 \pi^2 \left[ \left( \frac{\partial \Phi^{\text{dip}(1)}}{\partial r} \right)^2 + \left( \frac{\partial \Phi^{\text{dip}(1)}}{\partial t} \right)^2 \frac{1}{f^2} \right] B_{l'm'}
\end{equation}

\begin{equation}
B_{l' m'} = -\int d\Omega \, Y_{l'm'}^* \left\{ \left( \frac{\partial G}{\partial \theta} \right)^2 + \csc^2 \theta \left( \frac{\partial G}{\partial \phi} \right)^2 \right\}
\end{equation}

\begin{equation}
C_{l'm'} = -Y_{l' m'}^* \, \delta^2(\mathbf{0})
\end{equation}

\begin{equation}
Q^{r r} = 8 \pi \int d\Omega \, T^{\mu \nu} \mathcal{Y}_{\mu \nu(3)}^{*} = -\frac{f^2 \, \Phi^{\text{dip}(1)\, 2}}{r^2} C_{l'm'} + 16 \pi^2 \left[ f^2 \, \left( \frac{\partial \Phi^{\text{dip}(1)}}{\partial r} \right)^2 + \left( \frac{\partial \Phi^{\text{dip}(1)}}{\partial t} \right)^2  \right] B_{l'm'}
\end{equation}

\begin{equation}
Q^r = \frac{8 \pi r^2}{\lambda'} \int d\Omega \, T^{\mu \nu} \mathcal{Y}_{\mu \nu (6)}^* = -\Phi^{\text{dip}(1)} \frac{\partial \Phi^{\text{dip}(1)}}{\partial r} \frac{8 \pi f D_{l' m'}}{\lambda'}
\end{equation}

\begin{equation}
D_{l'm'} = \lambda' - 4 \pi \csc \theta_0 \sum_{l m} \frac{Y_{l m}^*}{\lambda'} \left\{ \dot{Y}_{l m} \dot{Y}_{l' m'}^* + \csc^2 \theta_0 Y_{l m}^{\prime} {Y}_{l' m'}^{\prime} \right\}
\end{equation}

\begin{equation}
Q^b = 8 \pi r^2 \int d\Omega \, T^{\mu \nu} \mathcal{Y}_{\mu \nu (4)}^* = 2 \frac{\Phi^{\text{dip}(1) \, 2}}{r^2} C_{l'm'} \, 
\end{equation}

\begin{equation}
Q^{\#} = \frac{32 \pi l' r^2}{\lambda' \mu'} \int d\Omega \, T^{\mu \nu} \mathcal{Y}_{\mu \nu (10)}^* = \frac{64  \pi^2 l' R_{l'm'} }{\lambda' \mu'} \left\{ f\left( \frac{\partial \Phi^{\text{dip}(1)}}{\partial r} \right)^2 - \frac{1}{f} \left( \frac{\partial \Phi^{\text{dip}(1)}}{\partial t} \right)^2 \right\}
\end{equation}

\begin{equation}
R_{l'm'} = 2\int d\Omega \Bigg[ \left( \frac{\partial G}{\partial \theta} \right)^2 + \csc^2 \theta \left( \frac{\partial G}{\partial \phi} \right)^2 \Bigg] \left[ \dot{Y}'_{l'm'} - \cot \theta Y'_{l'm'} \right] + 2 \frac{\partial G}{\partial \theta} \frac{\partial G}{\partial \phi} \csc^2 \theta \left[ \ddot{Y}_{l'm'} + \frac{1}{2} \lambda' \right]
\end{equation}

\section{Decomposition Using Spherical Harmonics in Schwarzschild Spacetime}\label{Decomposition Using Spherical Harmonics in Schwarzschild Spacetime}
    \noindent In this work we use the notation in[cite Karl's thesis and some papers doing the formalism for perturbation theory] for the Schwarzschild spacetime, where the metric that is expressed in spherical coordinate variables as 
    \begin{equation}
        ds^2 = -f(r)dt^2 +\frac{1}{f(r)}dr^2+r^2\left(d\theta^2 + \sin^2{\theta}d\phi^2\right)  
    \end{equation}
    where 
    \begin{equation}
        f(r) = 1-\frac{2M}{r}
    \end{equation}
    in geometric units, is factored as $U\times S_2$ as 
    \begin{equation}
        ds^2 = g_{ab}(x^1)dx^a dx^b + (x^1)^2\Omega_{AB}dx^A dx^B
    \end{equation}
    with indices $a,b = 0,1$ that correspond to the coordinate variables $t$ and $r$ parameterising the manifold $U$ and $A,B = 2,3$ correspond to $\theta$ and $\phi$ parameterising the sphere $S_2$. The notation for covariant derivative of a tensor $T$ defined on either the manifold $U$: $\tensor{T}{^{a_1}^{...}^{a_n}_{b_1}_{...}_{b_m}}$, or the sphere $S_2$: $\tensor{T}{^{A_1}^{...}^{A_n}_{B_1}_{...}_{B_m}}$ is 
    \begin{equation}
        \begin{aligned}
            &\nabla_C \tensor{T}{^{A_1}^{...}^{A_n}_{B_1}_{...}_{B_m}} = \tensor{T}{^{A_1}^{...}^{A_n}_{B_1}_{...}_{B_m}_{\mid C}} \\
            &\nabla_c \tensor{T}{^{a_1}^{...}^{a_n}_{b_1}_{...}_{b_m}} = \tensor{T}{^{a_1}^{...}^{a_n}_{b_1}_{...}_{b_m}_{:c}}
        \end{aligned}
    \end{equation}
    where $\nabla_C$ is the covariant derivative associated with the metric $\Omega_{AB}$ and $\nabla_c$ is the one that corresponds to $g_{ab}$; and for a tensor $\tensor{T}{^{\mu_1}^{...}^{\mu_n}_{\nu_1}_{...}_{\nu_m}}$, or the sphere $S_2$ defined on the whole spacetime $U\times S_2$ 
    \begin{equation}
        \begin{aligned}
            \nabla_{\sigma} \tensor{T}{^{\mu_1}^{...}^{\mu_n}_{\nu_1}_{...}_{\nu_m}} = \tensor{T}{^{\mu_1}^{...}^{\mu_n}_{\nu_1}_{...}_{\nu_m}_{;\sigma}}
        \end{aligned}
    \end{equation}
    where analogously $\nabla_{\sigma}$ corresponds to the metric of the whole spacetime metric $\overline{g}_{\mu \nu}$. On the other hand, normal partial differentiation is indicated by "$,$". 
    \noindent Moreover, the Levi-Civita tensor can be defined for both submanifolds \(U\) and \(S_2\) as follows 
    \begin{equation}
        \begin{aligned}
            &\epsilon_{ab} = \sqrt{-g} \ \varepsilon_{ab} \\
            &\epsilon_{AB} = \sqrt{\Omega} \ \varepsilon_{AB}
        \end{aligned}
    \end{equation}
    where \(\varepsilon_{01} = \varepsilon_{23} = 1\), and they satisfy the following contraction identities 
    \begin{equation}
        \begin{aligned}
            &\epsilon_a^{\ \ b} \epsilon_{bc} = g_{ac} \\
            &\epsilon_A^{\ \ B} \epsilon_{BC} = -\Omega_{AC}.
        \end{aligned}
    \end{equation}

    \noindent Now, to solve the coupled set of equations, namely the Einstein's field equations and Maxwell's field equations, we decompose tensors using spherical harmonics: $Y^{lm}(\theta,\phi)$, vector spherical harmonics
    \begin{equation}
        \begin{aligned}
            &Z_A^{lm}(\theta,\phi) = Y^{lm}_{\mid A}(\theta,\phi) \\
            &X_A^{lm}(\theta,\phi) = \tensor{\epsilon}{_B^A} Y^{lm}_{\mid A}(\theta,\phi)
        \end{aligned}
    \end{equation}
    where $\epsilon_{AB}$ is the Levi-Civita tensor on $S_2$, and tensor spherical harmonics
    \begin{equation}
        \begin{aligned}
            &U_{AB}(\theta,\phi) = \Omega_{AB} Y^{lm}(\theta,\phi) \\
            &V_{AB}(\theta,\phi) = Y^{lm}_{\mid AB}+\frac{l(l+1)}{2}\Omega_{AB}Y^{lm}(\theta,\phi) \\
            &W_{AB}(\theta,\phi) = X^{lm}_{(A\mid B)};
        \end{aligned}
    \end{equation}
    which define scalar, vectors, and (0,2)-tensors on $S_2$.   

    \noindent An important note here which we are going to make use of when evaluating integrals, is that the vector harmonics defined in (D6) correspond to the usual ones defined in Euclidean space $\mathbb{R}_3$ by 
    \begin{equation}
        \begin{aligned}
            &\mathbf{Z}^{lm} = r\nabla Y^{lm}(\theta,\phi)\\
            &\mathbf{X}^{lm} = \mathbf{r}\times\nabla Y^{lm}(\theta,\phi)
        \end{aligned}
    \end{equation}
    albeit for the normalisation of the unit vectors on the sphere by the metric $\Omega_{AB}$. In other words, we have the same dot products
    \begin{equation}
        \begin{aligned}
            &\mathbf{Z}^{lm}\cdot\mathbf{Z}^{l'm'} =\Omega^{AB}Z_A^{lm} Z_B^{l'm'}\\
            &\mathbf{X}^{lm}\cdot\mathbf{X}^{l'm'} = \Omega^{AB}X_A^{lm} X_B^{l'm'}\\
            &\mathbf{Z}^{lm}\cdot\mathbf{X}^{l'm'} = \Omega^{AB}Z_A^{lm} X_B^{l'm'}.
        \end{aligned}
    \end{equation}
    The vector harmonics defined in (D6) can be expressed in terms of the basis vector functions $\mathcal{Y}^{l}_{jm}(\theta,\phi)$ of the tensor product representation $D^l\otimes D^1$ of the rotation group in three dimensions given by 
    \begin{equation}
        \begin{aligned}
            \mathcal{Y}^{l}_{jm}(\theta,\phi) = \sum_{r=-l}^{r=l}\sum_{q=-1}^{q=1} (l,r;1,q|jm)Y^{lr}(\theta,\phi)\mathbf{e}_q
        \end{aligned}
    \end{equation}
    where $\mathbf{e}_q$ are the spherical vectors given by 
    \begin{equation}
        \begin{aligned}
            &\mathbf{e}_{\pm 1}=\frac{\mp 1}{\sqrt{2}}\left(\mathbf{\hat{x}}\pm i \mathbf{\hat{y}}\right) \\
            &\mathbf{e}_0 = \mathbf{\hat{z}},
        \end{aligned}
    \end{equation}
    as 
    \begin{equation}
        \begin{aligned}
            &\mathbf{Z}^{lm} = (l+1)\sqrt{\frac{l}{2l+1}}\mathcal{Y}^l_{l-1,m} + l\sqrt{\frac{l+1}{2l+1}}\mathcal{Y}^l_{l+1,m} \\
            &\mathbf{X}^{lm} = i\sqrt{l(l+1)}\mathcal{Y}^l_{lm}
        \end{aligned}
    \end{equation}
    which can be expanded in the vector basis $\mathbf{e}_q$ with components of spherical harmonic functions multiplied by numerical factors using (D10).

\section{Useful Spherical Integrals}\label{Useful Spherical Integrals}
    \noindent Given a gravitational mode \( (l,m) \) in the Regge-Wheeler/Zerilli-Moncrief equation, the electromagnetic source term that corresponds to the harmonic component of the source term of the same angular number \((l,m)\) is the projection on regular, vector, and tensor spherical harmonics of quadratic terms that are composed of two angular harmonic numbers \((l_1,m_1)\) and \((l_2,m_2)\).  Depending on which electromagnetic degrees of freedom, either of odd parity or even parity - see the next appendix, the type of coupling is different. Such two different modes of coupling is expressed in terms of the following two integrals
    \begin{equation}
        \begin{aligned}
            &\int_{S_2} Z^{l_1m_1\; A} X^{l_2m_2}_A Y^{l_3m_3}d\Omega = A(l_1,m_1;l_2,m_2;l_3,m_3)\\
            &\int_{S_2} Y^{l_1m_1}Y^{l_2m_2}Y^{l_3m_3}d\Omega = B(l_1,m_1;l_2,m_2;l_3,m_3),
        \end{aligned}
    \end{equation}
    where the first one is the coupling that comes from mixing odd and even degrees of freedom and the second appears for terms involving only terms of the same parity.  

    \noindent To evaluate the first integral, we compute the contraction of the two vector harmonics  
    \begin{equation}
        \begin{aligned}
            &Z^{l_1m_1\; A} X^{l_2m_2}_A = \Bigg( \frac{i}{2}(l_1+1)\sqrt{\frac{(l_1-m_1-1)(l_1+m_1+1)(l_2-m_2+1)(l_2+m_2)}{(2 l_1-1)(2l_1+1)}} Y^{l_2,m_2-1} Y^{l_1-1,m_1+1} - \\
            &\frac{i}{2}(l_1+1)\sqrt{\frac{(l_1+m_1-1)(l_1+m_1)(l_2-m_2)(l_2+m_2+1)}{(2 l_1-1)(2l_1+1)}} Y^{l_2,m_2+1} Y^{l_1-1,m_1-1} + i m_2(l_1+1) \\
            &\sqrt{\frac{(l_1-m_1)(l_1+m_1)}{(2 l_1-1)(2l_1+1)}} Y^{l_2 m_2} Y^{l_1-1,m_1} \Bigg) + \Bigg( \frac{i}{2} l_1 \sqrt{\frac{(l_1+m_1+1)(l_1+m_1+2)(l_2-m_2+1)(l_2+m_2)}{(2 l_1+1)(2l_1+3)}} \\
            & Y^{l_2,m_2-1} Y^{l_1+1,m_1+1} - \frac{i}{2} l_1 \sqrt{\frac{(l_1-m_1+1)(l_1-m_1+2)(l_2+m_2+1)(l_2-m_2)}{(2 l_1+1)(2l_1+3)}} Y^{l_2,m_2+1} Y^{l_1+1,m_1-1} \\
            & - i m_2 l_1 \sqrt{\frac{(l_1-m_1+1)(l_1+m_1+1)}{(2l_1-1)(2l_1+1)}} Y^{l_2 m_2} Y^{l_1+1,m_1} \Bigg).
        \end{aligned}
    \end{equation}
    \noindent Thus, the integral expressed in terms of \(B(l_1,m_1;l_2,m_2;l_3,m_3)\) becomes 
    \begin{equation}
        \begin{aligned}
            &\int_{S_2} Z^{l_1m_1\; A} X^{l_2m_2}_A Y^{l_3m_3}d\Omega = \Bigg( \frac{i}{2}(l_1+1)\sqrt{\frac{(l_1-m_1-1)(l_1+m_1+1)(l_2-m_2+1)(l_2+m_2)}{(2 l_1-1)(2l_1+1)}} \\
            &B(l_2,m_2-1;l_1-1,m_1+1;l,m) - \frac{i}{2}(l_1+1)\sqrt{\frac{(l_1+m_1-1)(l_1+m_1)(l_2-m_2)(l_2+m_2+1)}{(2 l_1-1)(2l_1+1)}} \\
            &B(l_2,m_2+1;l_1-1,m_1-1;l,m) + i m_2(l_1+1) \sqrt{\frac{(l_1-m_1)(l_1+m_1)}{(2 l_1-1)(2l_1+1)}} B(l_2 m_2;l_1-1,m_1;l,m) \Bigg) + \\
            &\Bigg( \frac{i}{2} l_1 \sqrt{\frac{(l_1+m_1+1)(l_1+m_1+2)(l_2-m_2+1)(l_2+m_2)}{(2 l_1+1)(2l_1+3)}} B(l_2,m_2-1;l_1+1,m_1+1;l,m) - \frac{i}{2} l_1 \\
            &\sqrt{\frac{(l_1-m_1+1)(l_1-m_1+2)(l_2+m_2+1)(l_2-m_2)}{(2 l_1+1)(2l_1+3)}} B(l_2,m_2+1;l_1+1,m_1-1;l,m) \\
            & - i m_2 l_1 \sqrt{\frac{(l_1-m_1+1)(l_1+m_1+1)}{(2l_1-1)(2l_1+1)}} B(l_2 m_2;l_1+1,m_1;l,m) \Bigg).
        \end{aligned}
    \end{equation}

    \noindent To evaluate the required integrals we use the following identities: 
    \begin{equation}
        \begin{aligned}
            &X^{lm}_{[A|B]} = \frac{1}{2}l(l+1)Y^{lm}\epsilon_{AB} \\
            &X_A^{l_1 m_1} X^{l_2 m_2 A} = Z_A^{l_1 m_1} Z^{l_2 m_2 A} \\
            &X^{l_2 m_2 A} Z^{l_1 m_1}_A = - X^{l_1 m_1 A} Z^{l_2 m_2}_A
        \end{aligned}
    \end{equation}
    
    \noindent Now, we evaluated the following integrals which are needed in the re-casting of the electromagentic energy-momentum tensor: 

    \noindent I.
    \begin{equation}
        \begin{aligned}
            &\int Z^{l_1 m_1 A} Z^{l_2 m_2}_A Y^{l_3 m_3} d \Omega=\frac{1}{2} \int \Omega^{A B} \nabla_A \nabla_B\left(Y^{l_2 m_2} Y^{l_1 m_1}\right) Y^{l_3 m_3} d \Omega \\
            & +\frac{1}{2}\left[l_1\left(l_1+1\right)+l_2\left(l_2+1\right)\right] \int Y^{l_{m_1}} Y^{l_2 m_2} Y^{l_3 m_3} d \Omega \\
            = & \frac{1}{2}\left[l_1\left(l_1+1\right)+l_2\left(l_2+1\right)-l_3\left(l_3+1\right)\right] \int Y^{l_1 m_1} Y^{l_2 m_2} Y^{l_3 m_3} d \Omega \\
            = & \frac{1}{2}\left[l_1\left(l_1+1\right)+l_2\left(l_2+1\right)-l_3\left(l_3+1\right)\right] B(l_1, m_1 ; l_2, m_2 ; l_3,m_3). 
        \end{aligned}
    \end{equation}

    \noindent II.
    \begin{equation}
        \begin{aligned}
            & \int Z^{l_1 m_1 A} Z^{l_2 m_2 B} Z_{A | B}^{l_3 m_3} d \Omega \\
            &= -\frac{1}{2}\left[\int \nabla_A\left(Z^{l_1 m_1 A} Z^{l_2 m_2 B}\right) Z_B^{l_3 m_3} d \Omega+\int \nabla_B\left(Z^{l_1 m_1 A} Z^{l_2 m_2 B}\right) Z_A^{l_3 m_3} d \Omega\right] \\
            & =\frac{1}{2} l_1(l_1+1) \int Y^{l_1 m_1} Z^{l_2 m_2 A} Z_A^{l_3 m_3} d \Omega+\frac{1}{2} l_2(l_2+1) \int Y^{l_2 m_2} Z^{l_1 m_1 A} Z^{l_3 m_3}_A d\Omega \\
            &-\frac{1}{2}\int \left(Z^{l_1 m_1}_{A|B} Z^{l_2 m_2 B} Z^{l_3 m_3 A} + Z^{l_1 m_1 B} Z^{l_2 m_2}_{B \mid A} Z^{l_3 m_3 A} \right)d\Omega.
        \end{aligned}
    \end{equation}
    \noindent But, 
    \begin{equation}
        \begin{aligned}
            & \int \left(Z^{l_1 m_1}_{A|B} Z^{l_2 m_2 B} Z^{l_3 m_3 A} + Z^{l_1 m_1 B} Z^{l_2 m_2}_{B \mid A} Z^{l_3 m_3 A} \right)d\Omega \\
            & =\int\left(Z^{l_1 m_1}_{B\mid A} Z^{l_1 m_2 B}+Z^{l_1 m_1 B} Z_{B \mid A}^{l_2 m_2}\right) Z^{l_3 m_3 A} d \Omega \\
            & =\int \nabla_A\left(Z^{l_1 m_1 B} Z_B^{l_2 m_2}\right) Z^{l_3 m_3 A} d \Omega=l_3(l_3+1) \int Z^{l_1 m_1 B} Z_B^{l_2 m_2} Y^{l_3 m_3} d \Omega.
        \end{aligned}
    \end{equation}
    \noindent Thus, we have 
    \begin{equation}
        \begin{aligned}
            &\int Z^{l_1 m_1 A} Z^{l_2 m_2 B} Z_{A \mid B}^{l_3 m_3} d \Omega= \frac{1}{4} \Bigg( l_1(l_1+1)\Big[l_2(l_2+1)+l_3(l_3+1)-l_1(l_1+1)\Big] \\
            &+ l_2(l_2+1)\Big[l_1(l_1+1)+l_3(l_3+1)-l_2(l_2+1)\Big] - l_3(l_3+1) \Big[l_1(l_1+1)+l_2(l_2+1)-l_3(l_3+1)\Big]\Bigg) \\
            &B(l_1,m_1;l_2,m_2;l_3,m_3)  = \Lambda (l_1,l_2,l_3)B(l_1,m_1;l_2,m_2;l_3,m_3).
        \end{aligned}
    \end{equation}
    
    \noindent III. 
    \begin{equation}
        \int X^{l_1 m_1 A}X^{l_2 m_2 B}Z^{l_3 m_3}_{A\mid B}d\Omega
    \end{equation}
    To evaluate this, we use the following three identities:
    \begin{equation}
        \begin{aligned}
            \int X^{l_1 m_1 A}X^{l_2 m_2}_{B\mid A} Z^{l_3 m_3 B} d\Omega = -\int X^{l_1 m_1 A} X^{l_2 m_2 B} Z^{l_3 m_3}_{A\mid B} d\Omega,
        \end{aligned}
    \end{equation}
    \begin{equation}
        \begin{aligned}
            &\int X^{l_1 m_1 A}X^{l_2 m_2}_{A\mid B} Z^{l_3 m_3 B} d\Omega = 2 \int X^{l_1 m_1 A}X^{l_2 m_2}_{[A\mid B]} Z^{l_3 m_3 B} d\Omega + \\
            &\int X^{l_1 m_1 A}X^{l_2 m_2}_{B\mid A} Z^{l_3 m_3 B} d\Omega,
        \end{aligned}
    \end{equation}
    and
    \begin{equation}
        \begin{aligned}
            &\int \left(X^{l_1 m_1 A}X^{l_2 m_2}_{A\mid B} Z^{l_3 m_3 B}+X^{l_1 m_1}_{A\mid B}X^{l_2 m_2 B} Z^{l_3 m_3 B} \right)d\Omega \\
            &=\int \nabla_B \left(X^{l_1 m_1 A} X^{l_2 m_2}_A\right) Z^{l_3 m_3 B}d\Omega = l_3(l_3+1)\int X^{l_1 m_1 A}X^{l_2 m_2}_A Y^{l_3 m_3} d\Omega.
        \end{aligned}
    \end{equation}
    Now, using (D11) and (D12) we get
    \begin{equation}
        \begin{aligned}
            &\int \left(X^{l_1 m_1 A} X^{l_2 m_2}_{A\mid B} Z^{l_3 m_3 B}+X^{l_1 m_1}_{A\mid B} X^{l_2 m_2 A} Z^{l_3 m_3 B} \right)d\Omega \\
            &=2 \int \left( X^{l_1 m_1 A} X^{l_2 m_2}_{[A\mid B]} + X^{l_1 m_1}_{[A\mid B]}X^{l_2 m_2 A}\right) Z^{l_3 m_3 B} d\Omega \\
            &+\int \left(X^{l_1 m_1}_{B\mid A}X^{l_2 m_2 A} Z^{l_3 m_3 B} +X^{l_1 m_1 A}X^{l_2 m_2}_{B\mid A} Z^{l_3 m_3 B} \right)d\Omega, 
        \end{aligned}
    \end{equation}
    and using (D10) we get 
    \begin{equation}
        \begin{aligned}
            &\int \left(X^{l_1 m_1 A} X^{l_2 m_2}_{A\mid B} Z^{l_3 m_3 B}+X^{l_1 m_1}_{A\mid B} X^{l_2 m_2 A} Z^{l_3 m_3 B} \right)d\Omega =2 \int \left(X^{l_1 m_1}_{[A\mid B]}X^{l_2 m_2 A} + X^{l_1 m_1 A} X^{l_2 m_2}_{[A\mid B]}\right)Z^{l_3 m_3 B} d\Omega \\
            &- 2\int X^{l_1 m_1 A} X^{l_2 m_2 B} Z^{l_3 m_3}_{A\mid B}d\Omega.
        \end{aligned}
    \end{equation}
    Hence, using (D12) and (D14) we get  
    \begin{equation}
        \begin{aligned}
            &\int X^{l_1 m_1 A} X^{l_2 m_2 B} Z^{l_3 m_3}_{A\mid B}d\Omega = \int X^{l_1 m_1}_{[A\mid B]}X^{l_2 m_2 A} Z^{l_3 m_3 B} d\Omega + \int X^{l_1 m_1 A} X^{l_2 m_2}_{[A\mid B]} Z^{l_3 m_3 B} d\Omega \\
            &-\frac{1}{2}l_3(l_3+1)\int X^{l_1 m_1 A}X^{l_2 m_2}_A Y^{l_3 m_3}d\Omega \\
            &=\frac{1}{2}l_1(l_1+1)\int Y^{l_1 m_1}Z^{l_2 m_2}_A Z^{l_3 m_3 A}d\Omega + \frac{1}{2}l_2(l_2+1)\int Y^{l_2 m_2}Z^{l_1 m_1}_A Z^{l_3 m_3 A} d\Omega \\
            &-\frac{1}{2}l_3(l_3+1)\int Y^{l_3 m_3} Z^{l_1 m_1}_A Z^{l_2 m_2 A}d\Omega \\
            & = \frac{1}{4} \Bigg( l_1(l_1+1)\Big[l_2(l_2+1)+l_3(l_3+1)-l_1(l_1+1)\Big] + l_2(l_2+1)\Big[l_1(l_1+1)+l_3(l_3+1)-l_2(l_2+1)\Big] \\
            &- l_3(l_3+1) \Big[l_1(l_1+1)+l_2(l_2+1)-l_3(l_3+1)\Big]\Bigg) B(l_1,m_1;l_2,m_2;l_3,m_3) \\
            & = \Lambda (l_1,l_2,l_3) B(l_1,m_1;l_2,m_2;l_3,m_3).
        \end{aligned}
    \end{equation}

    \noindent IV. 
    \begin{equation}
        \int X^{l_1 m_1 A} Z^{l_2 m_2 B} Z^{l_3 m_3}_{A\mid B} d\Omega
    \end{equation}
    \noindent To evaluate this, we use the following identities
    \begin{equation}
        \begin{aligned}
            &\int X^{l_1 m_1}_{A\mid B}Z^{l_2 m_2 A} Z^{l_3 m_3 B}d\Omega = -\int X^{l_1 m_1}_A \left(Z^{l_2 m_2 A}_{\quad\quad\quad \mid B} Z^{l_3 m_3 B} + Z^{l_2 m_2 A} Z^{l_3 m_3B}_{\quad \quad \quad \mid B}\right) d\Omega \\
            &= - \int X^{l_1 m_1 A} Z^{l_2 m_2}_{A\mid B} Z^{l_3 m_3 B}d\Omega + l_3(l_3+1)\int X_A^{l_1 m_1} Z^{l_2 m_2 A} Y^{l_3 m_3} d\Omega \\
            &=\int X^{l_1 m_1 A} Z^{l_2 m_2 B} Z^{l_3 m_3}_{A\mid B} d\Omega + l_3(l_3+1)\int X_A^{l_1 m_1} Z^{l_2 m_2 A}Y^{l_3 m_3} d\Omega,
        \end{aligned}
    \end{equation}
    \begin{equation}
        \begin{aligned}
             &\int X^{l_1 m_1}_{B\mid A}Z^{l_2 m_2 A} Z^{l_3 m_3 B} d\Omega = l_2(l_2+1)\int X^{l_1 m_1}_B Z^{l_3 m_3 B} Y^{l_2 m_2}d\Omega \\
             &-\int X^{l_1 m_1 B} Z^{l_2 m_2 A} Z^{l_3 m_3}_{A\mid B}d\Omega.
        \end{aligned}
    \end{equation}
    Subtracting (D18) from (D17) we get 
    \begin{equation}
        \begin{aligned}
            &\int X^{l_1 m_1 A} Z^{l_2 m_2 B} Z^{l_3 m_3}_{A\mid B}d\Omega = \frac{1}{2}l_1(l_1+1)\int X^{l_2 m_2 A} Z^{l_3 m_3}_A Y^{l_1 m_1}d\Omega \\
            &+\frac{1}{2}l_2(l_2+1)\int X^{l_1 m_1 A} Z^{l_3 m_3}_A Y^{l_2 m_2}d\Omega - \frac{1}{2}l_3(l_3+1)\int X^{l_1 m_1 A} Z^{l_2 m_2 A} Y^{l_3 m_3} d\Omega \\
            &=\frac{1}{2}l_1(l_1+1)A(l_3,m_3;l_2,m_2;l_1,m_1) + \frac{1}{2}l_2(l_2+1)A(l_3,m_3;l_1,m_1;l_2,m_2) \\
            &-\frac{1}{2}l_3(l_3+1)A(l_2,m_2;l_1,m_1;l_3,m_3).
        \end{aligned}
    \end{equation}

    \noindent V. 
    \begin{equation}
        \begin{aligned}
            &\int Z^{l_1 m_1 A} Z^{l_2 m_2 B} X^{l_3 m_3}_{(A\mid B)}d\Omega = \frac{1}{2} \Bigg[\int Z^{l_1 m_1 A} Z^{l_2 m_2 B} X^{l_3 m_3}_{A\mid B}d\Omega + \int Z^{l_1 m_1 A} Z^{l_2 m_2 B} X^{l_3 m_3}_{B\mid A}d\Omega\Bigg] \\
            &=\frac{1}{2}l_1(l_1+1)\int X_A^{l_3 m_3} Z^{l_2 m_2 A} Y^{l_3 m_3} d\Omega + \frac{1}{2} l_2(l_2+1) \int X_A^{l_3 m_3} Z^{l_1 m_1 A} Y^{l_2 m_2} d\Omega \\
            &=\frac{1}{2}l_1(l_1+1)A(l_2,m_2;l_3,m_3;l_1,m_1) + \frac{1}{2} l_2(l_2+1) A(l_1,m_1;l_3,m_3;l_2,m_2).
        \end{aligned}
    \end{equation}

    \noindent VI. 
    \begin{equation}
        \begin{aligned}
            &\int Z^{l_1 m_1 A} X^{l_2 m_2 B} X^{l_3 m_3}_{(A\mid B)}d\Omega = \int Z^{l_1 m_1 A} X^{l_2 m_2 B} \left( X^{l_3 m_3}_{A\mid B} - X^{l_3 m_3}_{[A\mid B]} \right) d\Omega \\
            &= \frac{1}{4} \Bigg[l_1(l_1+1)\Big(l_2(l_2+1)+l_3(l_3+1)-l_1(l_1+1)\Big) -l_2(l_2+1)\Big(l_1(l_1+1)+l_3(l_3+1)-l_2(l_2+1)\Big)\Bigg]\\
            &B(l_1,m_1;l_2,m_2;l_3,m_3).
        \end{aligned}
    \end{equation}

    \noindent VII. 
    \begin{equation}
        \begin{aligned}
            &\int X^{l_1 m_1 A} X^{l_2 m_2 B} X^{l_3 m_3}_{(A\mid B)} d\Omega = \frac{1}{2}\int Z^{l_1 m_1 A} X^{l_2 m_2 B} Z^{l_3 m_3}_{A\mid B}d\Omega + \frac{1}{2}\int X^{l_1 m_1 A} Z^{l_2 m_2 B} Z^{l_3 m_3}_{A\mid B} d\Omega \\
            &= \frac{1}{2} l_1(l_1+1)\int X^{l_2 m_2 A} Z^{l_3 m_3}_A Y^{l_1 m_1} d\Omega + \frac{1}{2} l_2(l_2+1)\int X^{l_1 m_1 A} Z^{l_3 m_3}_A Y^{l_2 m_2} d\Omega \\
            & - \frac{1}{2} l_3(l_3+1)\int X^{l_1 m_1 A} Z^{l_2 m_2}_A Y^{l_3 m_3} d\Omega - \frac{1}{2} l_3(l_3+1)\int X^{l_2 m_2 A} Z^{l_1 m_1}_A Y^{l_3 m_3} d\Omega \\
            &= \frac{1}{2} l_1(l_1+1) A(l_3,m_3;l_2,m_2;l_1,m_1)  + \frac{1}{2} l_2(l_2+1) A(l_3,m_3;l_1,m_1;l_2,m_2). 
        \end{aligned}
    \end{equation}   

\section{The Spherical Decomposition of the Electromagnetic Energy-Momentum Tensor}\label{The Spherical Decomposition of the Electromagnetic Energy-Momentum Tensor}
    \noindent The four covariant components of  the vector potential $A_{\mu}(x)$ are divided as $A_a(x)$ and $A_A(x)$ where $a=0,1$ and $A=2,3$, such that each are decomposed in the following manner
    \begin{equation}
        \begin{aligned}
            &A_a(x) =\sum_{l=0}^{\infty} \sum_{m=-l}^{m=l} g^{lm}_a(t,r) Y^{lm}(\theta,\phi) \\
            &A_A(x) = \sum_{l=0}^{\infty} \sum_{m=-l}^{m=l} \left[h^{lm}(t,r) Z_A^{lm}(\theta,\phi) + k^{lm}(t,r) X_A^{lm}(\theta,\phi)\right],
        \end{aligned}
    \end{equation} 
    and the four components of the current $J_{\mu}(x)$ can be decomposed similarly as 
    \begin{equation}
        \begin{aligned}
            &J_a(x) =\sum_{l=0}^{\infty} \sum_{m=-l}^{m=l} j^{lm}_a(t,r) Y^{lm}(\theta,\phi) \\
            &J_A(x) = \sum_{l=0}^{\infty} \sum_{m=-l}^{m=l} \left[j_3^{lm}(t,r) Z_A^{lm} (\theta,\phi) + j_4^{lm}(t,r) X_A^{lm}(\theta,\phi)\right].
        \end{aligned}
    \end{equation}

    \noindent The Maxwell-Stress tensor is given by 
    \begin{equation}
        \begin{aligned}
            \tensor{F}{_{\mu}_{\nu}} = A_{\nu,\mu}-A_{\mu,\nu} = \nabla_{\mu}A_{\nu}-\nabla_{\nu}A_{\mu}.
        \end{aligned}
    \end{equation}
    It then gives us the following decomposition: two scalars 
    \begin{equation}
        \tensor{F}{_a_b} = 2 \sum_{lm} g^{lm}_{[a,b]} Y^{lm},
    \end{equation}
    two vectors given by 
    \begin{equation}
        F_{aA} = \sum \Big[\left(h^{lm}_{,a}-g^{lm}_a\right)Z^{lm}_A+k^{lm}_{,a}X^{lm}_A\Big],
    \end{equation}
    a $(0,2)$-tensor 
    \begin{equation}
        F_{AB} = \sum_{lm} 2 k^{lm} X^{lm}_{[B|A]}.
    \end{equation}
    We will express the electromagnetic tensor \(F_{\mu \nu}\) and the energy-momentum tensor \(T_{\mu \nu}\) in terms of the following two quantities 
    \begin{equation}
        \begin{aligned}
            &{}_e\Phi^{lm}(t,r) = \frac{1}{\lambda_l}r^2 \left(g^{lm}_{1,0}(t,r)-g^{lm}_{0,1}(t,r)\right)\\
            &{}_o\Phi^{lm}(t,r) = k^{lm}(t,r)
        \end{aligned}
    \end{equation}
    and \(\lambda_l = l(l+1)\), where then we can write 
    \begin{equation}
        g^{lm}_{[a,b]} = -\frac{\lambda_l}{2r^2} \ {}_e\Phi^{lm}(t,r)\epsilon_{ab}.
    \end{equation}
    Moreover, from Maxwell's equation we get the following relation
    \begin{equation}
        g^{lm}_a-h^{lm}_{,a} = \frac{1}{\lambda_l}r^2 j_a^{lm} - \frac{1}{f} \ {}_e\Phi_{,a}^{lm}(t,r)
    \end{equation}
    where \(f=1-\frac{2M}{r}\), which we define as \({}_eK^{lm}_a(t,r)\). \\
    
    \noindent Thus, the electromagnetic tensor is expressed as 
    \begin{equation}
        \begin{aligned}
            &F_{ab} = -\frac{\lambda_l}{r^2} \ {}_e\Phi^{lm} \epsilon_{ab} Y^{lm}\\
            &F_{aA} = {}_eK^{lm}_a Z^{lm}_A + {}_o\Phi^{lm} X^{lm}_A \\
            &F_{AB} = - \lambda_l \ {}_o\Phi^{lm} \epsilon_{AB} Y^{lm}.
        \end{aligned}
    \end{equation}

    \noindent The energy-momentum tensor decomposes into: three scalars 
    \begin{equation}
        \begin{aligned}
            &T_{ab} = \sum_{l,m}\sum_{l'm'} \Bigg\{ -\frac{\lambda_l \lambda_{l'}}{2r^4}g_{ab} \Big({}_e\Phi^{lm} \ {}_e\Phi^{l'm'} + {}_o\Phi^{lm} \ {}_o\Phi^{l'm'} \Big) Y^{lm} Y^{l'm'} + \frac{1}{r^2} \Bigg({}_o\Phi^{l'm'}_{,b} \ {}_eK^{lm}_a - {}_o\Phi^{lm}_{,a} \ {}_eK^{l'm'}_b \\
            &- \frac{1}{2}g_{ab}\Big( {}_o\Phi^{l'm'}_{,c} \ {}_eK^{lm c} - {}_o\Phi^{lm}_{,c} \ {}_eK^{l'm' c}\Big)\Bigg) Z^{lm}_A X^{l'm' A} + \frac{1}{r^2} \Bigg({}_eK^{lm}_a \ {}_eK^{l'm'}_b + {}_o\Phi^{lm}_{,a} \ {}_o\Phi^{l'm'}_{,b} - \frac{1}{2}g_{ab} \\
            &\Big({}_eK^{lm}_c \ {}_eK^{l'm' c} + {}_o\Phi^{lm}_{,c} \ {}_o\Phi^{l'm' ,c }\Big)\Bigg) Z^{lm}_A Z^{l'm' A} \Bigg\},
        \end{aligned}
    \end{equation}
    two vectors 
    \begin{equation}
        \begin{aligned}
            &T_{aA} = \sum_{l,m}\sum_{l'm'} \frac{\lambda_l}{r^2} \Bigg\{ \Big({}_e\Phi^{lm} \epsilon_{ab} \ {}_eK^{l'm' b} + {}_o\Phi^{l'm'}_{,a} \ {}_o\Phi^{lm}\Big) Y^{lm} Z^{l'm'}_A + \Big({}_e\Phi^{lm} \epsilon_{ab} \ {}_o\Phi^{l'm' b} - {}_o\Phi^{lm} \ {}_eK^{l'm'}_a\Big) \\
            &Y^{lm} X^{l'm'}_A \Bigg\},
        \end{aligned}
    \end{equation}
    and a (0,2)-tensor 
    \begin{equation}
        \begin{aligned}
            &T_{AB} = \sum_{l,m}\sum_{l'm'} \Bigg\{ \frac{\lambda_l \lambda_{l'}}{2 r^2} \Omega_{AB} \Big({}_e\Phi^{lm} \ {}_e\Phi^{l'm'} + {}_o\Phi^{lm} \ {}_o\Phi^{l'm'}\Big) Y^{lm} Y^{l'm'} + {}_eK^{lm}_a \ {}_eK^{l'm' a} Z^{lm}_A Z^{l'm'}_B \\
            &+ {}_eK^{lm}_a \ {}_o\Phi^{l'm' ,a} \Big(Z^{lm}_A X^{l'm'}_B + Z^{lm}_B X^{l'm'}_A\Big) + {}_o\Phi^{lm}_{,a} \ {}_o\Phi^{l'm' ,a} X^{lm}_A X^{l'm'}_B - \frac{1}{2r^2}\Omega_{AB}\Big({}_eK^{lm}_a \ {}_o\Phi^{l'm' ,a} - {}_eK^{l'm'}_a \ \\
            &{}_o\Phi^{lm ,a}\Big) Z^{lm}_C X^{l'm' C} - \frac{1}{2r^2}\Omega_{AB} \Big( {}_eK^{lm}_a \ {}_eK^{l'm' a} + {}_o\Phi^{lm}_{,a} \ {}_o\Phi^{l'm' ,a} \Big) Z^{lm}_C Z^{l'm' C} \Bigg\}
        \end{aligned}
    \end{equation}

\section{Electromagnetic Source Terms for the Regge-Wheeler and Zerilli-Moncrief Equations}\label{Electromagnetic Source Terms for the Regge-Wheeler and Zerilli-Moncrief Equations}
    \noindent We have two source terms coming from the electromagentic energy-momentum tensor, one for the even gravitational perturbation scalar ${}_e\Psi$ and one for the odd gravitational perturbation scalar ${}_o\Psi$.

    \noindent The source terms of both differential equations are written in terms of the decomposition of the electromagnetic energy-momentum tensor into three scalars 
    \begin{equation}
        \begin{aligned}
            &Q_{ab}^{lm} = 8\pi \int T_{ab} \overline{Y^{lm}}d\Omega \\
            & = (-1)^m 8\pi \sum_{l_1 m_1}\sum_{l_2 m_2} \Bigg\{ \Bigg[ -\frac{1}{2}g_{ab} \Bigg(\frac{\lambda_{l_1} \lambda_{l_2}}{r^4}\Big( {}_e\Phi^{l_1 m_1} \ {}_e\Phi^{l_2 m_2} + {}_o\Phi^{l_1 m_1} \ {}_o\Phi^{l_2 m_2}\Big) + \frac{1}{2}\Big(\lambda_{l_1} + \lambda_{l_2} -\lambda_l\Big)\\
            &\Big( {}_eK^{l_1 m_1}_c \ {}_eK^{l_2 m_2 c} +  {}_o\Phi^{l_1 m_1}_{,c} \ {}_o\Phi^{l_2 m_2 ,c} \Bigg) + \frac{\lambda_{l_1}+\lambda_{l_2}-\lambda_l}{2r^2} \Big( {}_eK^{l_1 m_1}_a \ {}_eK^{l_2 m_2}_b + {}_o\Phi^{l_1 m_1}_{,a} \ {}_o\Phi^{l_2 m_2}_{,b}\Big) \Bigg] \\
            &B(l_1,m_1;l_2,m_2;l,-m) + \frac{1}{r^2} \Bigg[ {}_o\Phi^{l_2 m_2}_{,b} \ {}_eK^{l_1 m_1}_a - {}_o\Phi^{l_1 m_1}_{,a} \ {}_eK^{l_2 m_2}_b - \frac{1}{2}g_{ab}\Big( {}_o\Phi^{l_2 m_2}_{,c} \ {}_eK^{l_1 m_1 c} - {}_o\Phi^{l_1 m_1}_{,c} \ {}_eK^{l_2 m_2 c} \Big) \Bigg] \\
            &A(l_1,m_1;l_2,m_2;l,-m)\Bigg\}, 
        \end{aligned}
    \end{equation}
    two vectors
    \begin{equation}
        \begin{aligned}
            &Q^{lm}_a = \frac{16 \pi r^2}{\lambda_l} \int T_{aA} \overline{Z^{lm A}} d\Omega \\
            & = (-1)^m \frac{16 \pi r^2}{\lambda_l}\sum_{l_1 m_1}\sum_{l_2 m_2} \frac{\lambda_{l_1}}{r^2} \Bigg\{ \frac{1}{2}\Big( \lambda_{l_2} + \lambda_l - \lambda_{l_1} \Big) \Big( {}_e\Phi^{l_1 m_1} \epsilon_{ab} \ {}_eK^{l_2 m_2 b} + {}_o\Phi^{l_2 m_2}_{,a} \ {}_o\Phi^{l_1 m_1} \Big)B(l_1,m_1;l_2,m_2;l_3,m_3) \\
            &+\Big({}_e\Phi^{l_1 m_1} \epsilon_{ab} \ {}_o\Phi^{l_2 m_2 ,b} - {}_o\Phi^{l_1 m_1} \ {}_eK^{l_2 m_2}_a\Big) A(l,-m;l_2,m_2;l_1,m_1)\Bigg\},
        \end{aligned}
    \end{equation}
    \begin{equation}
        \begin{aligned}
            &P^{lm}_a = \frac{16 \pi r^2}{\lambda_l} \int T_{aA} \overline{X^{lm A}} d\Omega \\
            &= (-1)^m \frac{16 \pi r^2}{\lambda_l}\sum_{l_1 m_1}\sum_{l_2 m_2} \frac{\lambda_{l_1}}{r^2} \Bigg\{ \Big( {}_e\Phi^{l_1 m_1} \epsilon_{ab} \ {}_eK^{l_2 m_2 b} + {}_o\Phi^{l_2 m_2}_{,a} \ {}_o\Phi^{l_1 m_1} \Big) A(l_2,m_2;l,-m;l_1,m_1) \\
            &+ \frac{1}{2}\Big( \lambda_{l_2} + \lambda_l - \lambda_{l_1} \Big) \Big({}_e\Phi^{l_1 m_1} \epsilon_{ab} \ {}_o\Phi^{l_2 m_2 ,b} - {}_o\Phi^{l_1 m_1} \ {}_eK^{l_2 m_2}_a\Big) B(l_1,m_1;l_2,m_2;l,-m)\Bigg\};
        \end{aligned}
    \end{equation}
    and three tensors 
    \begin{equation}
        \begin{aligned}
            &Q^{lm \flat} = 8\pi r^2 \int T_{AB} \overline{U^{lm AB}} d\Omega \\
            &= (-1)^m 8\pi r^2 \sum_{l_1 m_1}\sum_{l_2 m_2} \Bigg\{ \Bigg[ \frac{\lambda_{l_1} \lambda_{l_2}}{r^2} \Big( {}_e\Phi^{l_1 m_1} \ {}_e\Phi^{l_2 m_2} + {}_o\Phi^{l_1 m_1} \ {}_o\Phi^{l_2 m_2}\Big) +\frac{1}{2} \Big(\lambda_{l_1}+\lambda_{l_2}-\lambda_l\Big)\Big(1-\frac{1}{r^2}\Big) \\
            &\Big({}_eK^{l_1 m_1}_a \ {}_eK^{l_2 m_2 a} + {}_o\Phi^{l_1 m_1}_{,a} \ {}_o\Phi^{l_2 m_2 ,a} \Big) \Bigg] B(l_1,m_1;l_2,m_2;l,-m) + \Bigg[ 2 {}_eK^{l_1 m_1}_a \ {}_o\Phi^{l_2 m_2 ,a} - \frac{1}{r^2}\Big( {}_eK^{l_1 m_1}_a \ {}_o\Phi^{l_2 m_2 ,a} - \\
            &{}_eK^{l_2 m_2}_a \ {}_o\Phi^{l_1 m_1 ,a} \Big)\Bigg] A(l_1,m_1;l_2,m_2;l,-m)\Bigg\}
        \end{aligned}
    \end{equation}
    \begin{equation}
        \begin{aligned}
            &Q^{lm \sharp} = 32\pi \frac{(l-2)!}{(l+2)!}r^4 \int T_{AB} \overline{V^{lm AB}} d\Omega \\
            &= (-1)^m 32\pi \frac{(l-2)!}{(l+2)!}r^4 \sum_{l_1 m_1}\sum_{l_2 m_2} \Bigg\{ \Big( \Lambda(l_1,l_2,l) + \frac{1}{4}\lambda_l(\lambda_{l_1}+\lambda_{l_2}-\lambda_l) \Big)\Big( {}_eK^{l_1 m_1}_a \ {}_eK^{l_2 m_2 a} + {}_o\Phi^{l_2 m_2}_{,a} \ {}_o\Phi^{l_1 m_1 ,a} \Big) \\
            &B(l_1,m_1;l_2,m_2;l,-m) + {}_eK^{l_1 m_1}_a \ {}_o\Phi^{l_2 m_2 ,a} \Big(\lambda_{l_1} A(l,-m;l_2,m_2;l_1,m_1) + \lambda_{l_2} A(l,-m;l_1,m_1;l_2,m_2)\Big) \Bigg\},
        \end{aligned}
    \end{equation}
    \begin{equation}
        \begin{aligned}
            &P = 16\pi \frac{(l-2)!}{(l+2)!}r^4 \int T_{AB} \overline{W^{lm AB}} d\Omega \\
            &= (-1)^m 16\pi \frac{(l-2)!}{(l+2)!}r^4 \sum_{l_1 m_1}\sum_{l_2 m_2} \Bigg\{ \frac{1}{2} \Big( {}_eK^{l_1 m_1}_a \ {}_eK^{l_2 m_2 a} - {}_o\Phi^{l_2 m_2}_{,a} \ {}_o\Phi^{l_1 m_1 ,a} \Big) \Big(\lambda_{l_1} A(l_2,m_2;l,-m;l_1,m_1) \\
            &+ \lambda_{l_2} A(l_1,m_1;l,-m;l_2,m_2)\Big) + 2 \Big( \Lambda(l_1,l,l_2)-\frac{1}{4}\lambda_l(\lambda_{l_1}+\lambda_{l_2}-\lambda_l) \Big) {}_eK^{l_1 m_1}_a \ {}_o\Phi^{l_2 m_2 ,a} B(l_1,m_1;l_2,m_2;l,-m).       
        \end{aligned}
    \end{equation}

    \noindent The source term for the Zerilli-Moncrief equation $S^{lm}_{ZM}$ for even metric perturbations is given by 
    \begin{equation}
        \begin{aligned}
            &S^{lm}_{ZM} = \frac{2}{\Lambda} r^{,a} Q^{lm}_a - \frac{1}{r} Q^{lm \sharp} + \frac{r^2}{(\lambda + 1)\Lambda} \Big(-(Q^{lm a}_{\quad \ \ a})_{,b} r^{,b} + \frac{6M}{r^2 \Lambda} Q^{lm}_{ab} r^{,a}r^{,b} + \frac{f}{r} Q^{lm \flat} \\
            &+ \frac{1}{r\Lambda} \left( \lambda (\lambda + 1) +\frac{3M}{r}(2 \lambda -3) + \frac{21 M^2}{r^2} \right) Q^{lm a}_{\quad \ \ a} \Bigg), 
        \end{aligned}
    \end{equation}
    and the source term for the Regge-Wheeler equation $S^{lm}_{RW}$ for odd metric perturbations is given by 
    \begin{equation}
        S^{lm}_{RW} = \frac{2}{r^2} \left( 1 - \frac{3M}{r} \right) P^{lm} + \frac{r^{,a}}{r} \left(P^{lm}_a - P^{lm}_{,a} \right), 
    \end{equation}
    where \(\lambda = (l+2)(l-1)/2\) and \(\Lambda = \lambda + 3M/r\).

\end{widetext}
\bibliography{main}

\end{document}